\documentclass[
aps, 
prl,
preprintnumbers,
twocolumn,
superscriptaddress,
nofootinbib,
floatfix,
10pt
]{revtex4-2}

\usepackage[percent]{overpic}
\usepackage{booktabs}
\usepackage{graphicx}
\usepackage{dcolumn}
\usepackage{bm}
\usepackage{hyperref}
\usepackage[ulem=normalem]{changes} 
\usepackage{adjustbox}

\usepackage{comment}
\usepackage{amsmath}

\def\orcid#1{\kern .08em\href{https://orcid.org/#1}{\includegraphics[keepaspectratio,width=0.7em]{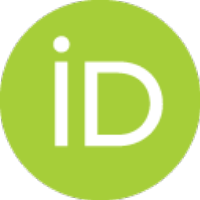}}}

\begin{document}

\title{Direct proton transfer on $^{46}$Ar supports the presence of a charge density bubble linked to a  novel nuclear structure below $^{48}$Ca}

\author{Daniele~Brugnara\orcid{0000-0002-8644-5355}}
\email{dbrugnara@lnl.infn.it}
\affiliation{Dipartimento di Fisica e Astronomia, Universit\`a di Padova, Padova, I-35131, Italy}
\affiliation{Istituto Nazionale di Fisica Nucleare, Laboratori Nazionali di Legnaro, Legnaro, I-35020, Italy}

\author{Andrea~Gottardo\orcid{0000-0002-0390-5767}}
\affiliation{Istituto Nazionale di Fisica Nucleare, Laboratori Nazionali di Legnaro, Legnaro, I-35020, Italy}

\author{Marlene~Assi\`e}
\affiliation{Université Paris-Saclay, CNRS/IN2P3, IJCLab, Orsay, F-91405, France}

\author{Carlo~Barbieri\orcid{0000-0001-8658-6927}}
\affiliation{Dipartimento di Fisica “Aldo Pontremoli,” Universit\`a degli Studi di Milano, Milano, I-20133, Italy}
\affiliation{INFN, Sezione di Milano, Milano, I-20133, Italy}

\author{Daniele~Mengoni\orcid{0000-0001-7219-2548}}
\affiliation{Dipartimento di Fisica e Astronomia, Universit\`a di Padova, Padova, I-35131, Italy}
\affiliation{INFN, Sezione di Padova, Padova, I-35131, Italy}

\author{Didier~Beaumel}
\affiliation{Université Paris-Saclay, CNRS/IN2P3, IJCLab, Orsay, F-91405, France}

\author{Stefano~Brolli\orcid{0009-0005-8290-8462}}
\affiliation{Dipartimento di Fisica “Aldo Pontremoli,” Universit\`a degli Studi di Milano, Milano, I-20133, Italy}
\affiliation{INFN, Sezione di Milano, Milano, I-20133, Italy}

\author{Simone~Bottoni\orcid{0000-0003-2249-4825}}
\affiliation{Dipartimento di Fisica “Aldo Pontremoli,” Universit\`a degli Studi di Milano, Milano, I-20133, Italy}
\affiliation{INFN, Sezione di Milano, Milano, I-20133, Italy}

\author{Emmanuel~Clement}
\affiliation{Grand Accélérateur National d’Ions Lourds, CEA/DRF-CNRS/IN2P3, Caen, F-14076, France}

\author{Gianluca~Colò\orcid{0000-0003-0819-1633}}
\affiliation{Dipartimento di Fisica “Aldo Pontremoli,” Universit\`a degli Studi di Milano, Milano, I-20133, Italy}
\affiliation{INFN, Sezione di Milano, Milano, I-20133, Italy}

\author{Freddy~Flavigny}
\affiliation{Normandie Université, ENSICAEN, UNICAEN, CNRS/IN2P3, LPC Caen, Caen, F-14000, France}

\author{Franco~Galtarossa}
\affiliation{Université Paris-Saclay, CNRS/IN2P3, IJCLab, Orsay, F-91405, France}
\affiliation{INFN, Sezione di Padova, Padova, I-35131, Italy}

\author{Valerian~Girard-Alcindor}
\affiliation{Université Paris-Saclay, CNRS/IN2P3, IJCLab, Orsay, F-91405, France}
\affiliation{Grand Accélérateur National d’Ions Lourds, CEA/DRF-CNRS/IN2P3, Caen, F-14076, France}

\author{Antoine~Lemasson}
\affiliation{Grand Accélérateur National d’Ions Lourds, CEA/DRF-CNRS/IN2P3, Caen, F-14076, France}

\author{Adrien~Matta}
\affiliation{Normandie Université, ENSICAEN, UNICAEN, CNRS/IN2P3, LPC Caen, Caen, F-14000, France}

\author{Diego~Ramos}
\affiliation{Grand Accélérateur National d’Ions Lourds, CEA/DRF-CNRS/IN2P3, Caen, F-14076, France}

\author{Vittorio~Som\`a\orcid{0000-0001-9386-4104}}
\affiliation{Université Paris-Saclay, IRFU, CEA, Gif-sur-Yvette, F-91191, France}

\author{Jose Javier Valiente-Dobón\orcid{0000-0002-8651-1957}}
\affiliation{Istituto Nazionale di Fisica Nucleare, Laboratori Nazionali di Legnaro, Legnaro, I-35020, Italy}

\author{Enrico~Vigezzi\orcid{0000-0002-2683-6442}}
\affiliation{INFN, Sezione di Milano, Milano, I-20133, Italy}

\author{Mathieu~Babo}
\affiliation{Université Paris-Saclay, CNRS/IN2P3, IJCLab, Orsay, F-91405, France}

\author{Diego~Barrientos}
\affiliation{CERN, Espl. des Particules 1, Meyrin, CH-1211, Switzerland}

\author{Dino~Bazzacco}
\affiliation{INFN, Sezione di Padova, Padova, I-35131, Italy}

\author{Piotr~Bednarczyk}
\affiliation{Polish Academy of Sciences, The Henryk Niewodniczański Institute of Nuclear Physics, Kraków, 31-342, Poland}

\author{Giovanna~Benzoni\orcid{0000-0002-7938-0338}}
\affiliation{INFN, Sezione di Milano, Milano, I-20133, Italy}

\author{Yorick~Blumenfeld}
\affiliation{Université Paris-Saclay, CNRS/IN2P3, IJCLab, Orsay, F-91405, France}

\author{Helen~Boston}
\affiliation{Oliver Lodge Laboratory, University of Liverpool, Liverpool, L69 7ZE, United Kingdom}

\author{Angela~Bracco\orcid{0000-0003-3370-4488}}
\affiliation{Dipartimento di Fisica “Aldo Pontremoli,” Universit\`a degli Studi di Milano, Milano, I-20133, Italy}
\affiliation{INFN, Sezione di Milano, Milano, I-20133, Italy}

\author{Bo~Cederwall}
\affiliation{Department of Physics, KTH Royal Institute of Technology, Stockholm, SE-10691, Sweden}

\author{Michal~Ciemala}
\affiliation{Polish Academy of Sciences, The Henryk Niewodniczański Institute of Nuclear Physics, Kraków, 31-342, Poland}

\author{Ushasi~Datta}
\affiliation{Nuclear Physics Division, Saha Institute of Nuclear Physics, Kolkata, 700064, India}
\affiliation{Homi Bhabha National Institute, Mumbai, 400094, India}

\author{Giacomo~de~Angelis}
\affiliation{Istituto Nazionale di Fisica Nucleare, Laboratori Nazionali di Legnaro, Legnaro, I-35020, Italy}

\author{Gilles~de~France}
\affiliation{Grand Accélérateur National d’Ions Lourds, CEA/DRF-CNRS/IN2P3, Caen, F-14076, France}

\author{César~Domingo-Pardo}
\affiliation{Instituto de Física Corpuscular, CSIC-Universidad de Valencia, Burjassot, Valencia,
E-46071, Spain}

\author{Jérémie~Dudouet}
\affiliation{Université de Lyon, Université Lyon-1, CNRS/IN2P3, UMR5822, IPNL, Villeurbanne Cedex, F-69622, France}

\author{Jose~Dueñas\orcid{0000-0003-2940-5135}}
\affiliation{Departamento de Ingeniería Eléctrica y Centro de Estudios Avanzados en Física, Matemáticas y Computación, Universidad de Huelva, 21007 Huelva, Spain}

\author{Beatriz~Fernandez~Dominguez}
\affiliation{Departamento de Física de Partículas, Universidade de Santiago de Compostela, Santiago de Compostela, 15782, Spain}

\author{Andres~Gadea}
\affiliation{Instituto de Física Corpuscular, CSIC-Universidad de Valencia, Burjassot, Valencia,
E-46071, Spain}

\author{Alain~Goasduff}
\affiliation{Istituto Nazionale di Fisica Nucleare, Laboratori Nazionali di Legnaro, Legnaro, I-35020, Italy}

\author{Vicente~González}
\affiliation{Departamento de Ingeniería Electrónica Universitat de Valencia, Burjassot, Valencia, E-46100, Spain}

\author{Fairouz~Hammache}
\affiliation{Université Paris-Saclay, CNRS/IN2P3, IJCLab, Orsay, F-91405, France}

\author{Laura~Harkness-Brennan}
\affiliation{Oliver Lodge Laboratory, University of Liverpool, Liverpool, L69 7ZE, United Kingdom}

\author{Herbert~Hess}
\affiliation{Institut für Kernphysik, Universität zu Köln, Köln, D-50937, Germany}

\author{Charles~Houarner}
\affiliation{Grand Accélérateur National d’Ions Lourds, CEA/DRF-CNRS/IN2P3, Caen, F-14076, France}

\author{Andrés~Illana \orcid{0000-0003-0274-3388}}
\affiliation{Grupo de Física Nuclear and IPARCOS, Universidad Complutense de Madrid, CEI Moncloa, E-28040, Spain}
\affiliation{Istituto Nazionale di Fisica Nucleare, Laboratori Nazionali di Legnaro, Legnaro, I-35020, Italy}

\author{Daniel~Judson}
\affiliation{Oliver Lodge Laboratory, University of Liverpool, Liverpool, L69 7ZE, United Kingdom}

\author{Andrea~Jungclaus}
\affiliation{Instituto de Estructura de la Materia, CSIC,Madrid, E-28006, Spain}

\author{Wolfram~Korten}
\affiliation{Université Paris-Saclay, IRFU, CEA, Gif-sur-Yvette, F-91191, France}

\author{Marc~Labiche}
\affiliation{STFC Daresbury Laboratory, Daresbury, Warrington, WA4 4AD, United Kingdom}

\author{Louis~Lalanne}
\affiliation{Université Paris-Saclay, CNRS/IN2P3, IJCLab, Orsay, F-91405, France}

\author{Silvia~Leoni\orcid{0000-0002-3691-0749}}
\affiliation{Dipartimento di Fisica “Aldo Pontremoli,” Universit\`a degli Studi di Milano, Milano, I-20133, Italy}
\affiliation{INFN, Sezione di Milano, Milano, I-20133, Italy}

\author{Cyril~Lenain}
\affiliation{Normandie Université, ENSICAEN, UNICAEN, CNRS/IN2P3, LPC Caen, Caen, F-14000, France}

\author{Sylvain~Leblond}
\affiliation{Université Paris-Saclay, IRFU, CEA, Gif-sur-Yvette, F-91191, France}

\author{Joa~Ljungvall}
\affiliation{Université Paris-Saclay, CNRS/IN2P3, IJCLab, Orsay, F-91405, France}
\affiliation{IPHC, CNRS, Université de Strasbourg, Strasbourg, F-67037, France}

\author{Ivano~Lombardo}
\affiliation{Dipartimento di Fisica e Astronomia, Università di Catania, Catania, I-95125, Italy}
\affiliation{INFN, Sezione di Catania, Catania, I-95125, Italy}

\author{Adam~Maj}
\affiliation{Polish Academy of Sciences, The Henryk Niewodniczański Institute of Nuclear Physics, Kraków, 31-342, Poland}

\author{Naomi~Marchini}
\affiliation{INFN, Sezione di Firenze, Sesto Fiorentino, Firenze, 50019, Italy}

\author{Roberto~Menegazzo}
\affiliation{INFN, Sezione di Padova, Padova, I-35131, Italy}

\author{Bénédicte~Million\orcid{0000-0001-7988-9911}}
\affiliation{INFN, Sezione di Milano, Milano, I-20133, Italy}

\author{Johan~Nyberg}
\affiliation{Department of Physics and Astronomy, Uppsala University, Uppsala, SE-75120, Sweden}

\author{Rosa~Maria~Pérez-Vidal}
\affiliation{Instituto de Física Corpuscular, CSIC-Universidad de Valencia, Valencia E-46980, Spain}

\author{Zsolt~Podolyak}
\affiliation{Department of Physics, University of Surrey, Guildford, GU2 7XH, United Kingdom}

\author{Alberto~Pullia\orcid{0000-0002-6393-747X}}
\affiliation{Dipartimento di Fisica “Aldo Pontremoli,” Universit\`a degli Studi di Milano, Milano, I-20133, Italy}
\affiliation{INFN, Sezione di Milano, Milano, I-20133, Italy}

\author{Begoña~Quintana}
\affiliation{Laboratorio de Radiaciones Ionizantes, Departamento de Física Fundamental, Universidad de Salamanca, Salamanca, E-37008, Spain}

\author{Francesco~Recchia}
\affiliation{Dipartimento di Fisica e Astronomia, Universit\`a di Padova, Padova, I-35131, Italy}
\affiliation{INFN, Sezione di Padova, Padova, I-35131, Italy}

\author{Peter~Reiter}
\affiliation{Institut für Kernphysik, Universität zu Köln, Köln, D-50937, Germany}

\author{Kseniia~Rezynkina}
\affiliation{Dipartimento di Fisica e Astronomia, Universit\`a di Padova, Padova, I-35131, Italy}

\author{Marco~Rocchini}
\affiliation{INFN, Sezione di Firenze, Sesto Fiorentino, Firenze, 50019, Italy}

\author{Frédéric~Saillant}
\affiliation{Grand Accélérateur National d’Ions Lourds, CEA/DRF-CNRS/IN2P3, Caen, F-14076, France}

\author{Marie-Delphine~Salsac}
\affiliation{Université Paris-Saclay, IRFU, CEA, Gif-sur-Yvette, F-91191, France}

\author{Angel~Miguel~Sanchez~Benitez}
\affiliation{Department of Integrated Sciences, Centro de Estudios Avanzados en Física, Matemáticas y Computación (CEAFMC), Universidad de Huelva, Huelva, 21071, Spain}

\author{Jennifer~Sanchez~Rojo}
\affiliation{TRIUMF, Vancouver, British Columbia, 4004, Canada}

\author{Enrique~Sanchis}
\affiliation{Instituto de Física Corpuscular, CSIC-Universidad de Valencia, Burjassot, Valencia,
E-46071, Spain}

\author{Menekşe~Şenyiğit}
\affiliation{Department of Physics, Ankara University, Besevler, Ankara, 06100, Turkey}

\author{Marco~Siciliano}
\affiliation{Université Paris-Saclay, IRFU, CEA, Gif-sur-Yvette, F-91191, France}
\affiliation{Physics Division, Argonne National Laboratory, Lemont, Illinois, 60439, United States}

\author{John~Simpson}
\affiliation{STFC Daresbury Laboratory, Daresbury, Warrington, WA4 4AD, United Kingdom}

\author{Dorottya~Sohler}
\affiliation{HUN-REN Institute for Nuclear Research, HUN-REN Atomki, Debrecen, 4001, Hungary}

\author{Christophe~Theisen}
\affiliation{Université Paris-Saclay, IRFU, CEA, Gif-sur-Yvette, F-91191, France}

\author{Irene~Zanon}
\affiliation{Istituto Nazionale di Fisica Nucleare, Laboratori Nazionali di Legnaro, Legnaro, I-35020, Italy}

\author{Magdalena~Zielinska}
\affiliation{Université Paris-Saclay, IRFU, CEA, Gif-sur-Yvette, F-91191, France}

\date{\today}

\begin{abstract}
The $^{46}$Ar($^3$He,d)$^{47}$K reaction was performed in inverse kinematics using a radioactive $^{46}$Ar beam produced by the SPIRAL1 facility at GANIL and a cryogenic $^{3}$He target. The AGATA-MUGAST-VAMOS setup allowed the coincident measurement of the $\gamma$ rays, deuterons and recoiling $^{47}$K isotopes produced by the reaction. The relative cross sections towards the proton-addition states in $^{47}$K point towards a depletion of the $\pi s_{1/2}$ shell.
The  experimental findings are in good agreement with  ab initio calculations, which predict that  $^{46}$Ar exhibits a charge density bubble associated with a pronounced proton closed-shell character.  
\end{abstract}

\maketitle

The short-range nature of the nuclear force leads to densely packed nucleons forming a strongly correlated quantum fluid and, in most cases, a nearly constant density inside the nucleus~\cite{DickBarb2004PPNP,Sick2007Correls}. 
Yet, nucleons can be interpreted as occupying energy levels organised in shells, leading to the emergence of configurations with remarkable stability, in analogy with electrons in noble gas atoms~\cite{Mayer1949SM,Caurier2005rmpSM}. 
Unravelling the interplay between collective fluid-like and single-particle aspects in unstable neutron-rich isotopes stands as a pivotal challenge in modern nuclear physics. 
This knowledge holds the potential to provide valuable insights not only into the understanding of nuclear structure but also into phenomena like the ``cold r-process'' responsible for the nucleosynthesis of elements during neutron star mergers, ultimately linked to the abundance of elements in the Universe~\cite{MUMPOWER201686,Siegel2022Nature_rProc}.

\begin{figure*}[htb!]%
\centering
\centering
\begin{overpic}[width=0.74\textwidth]{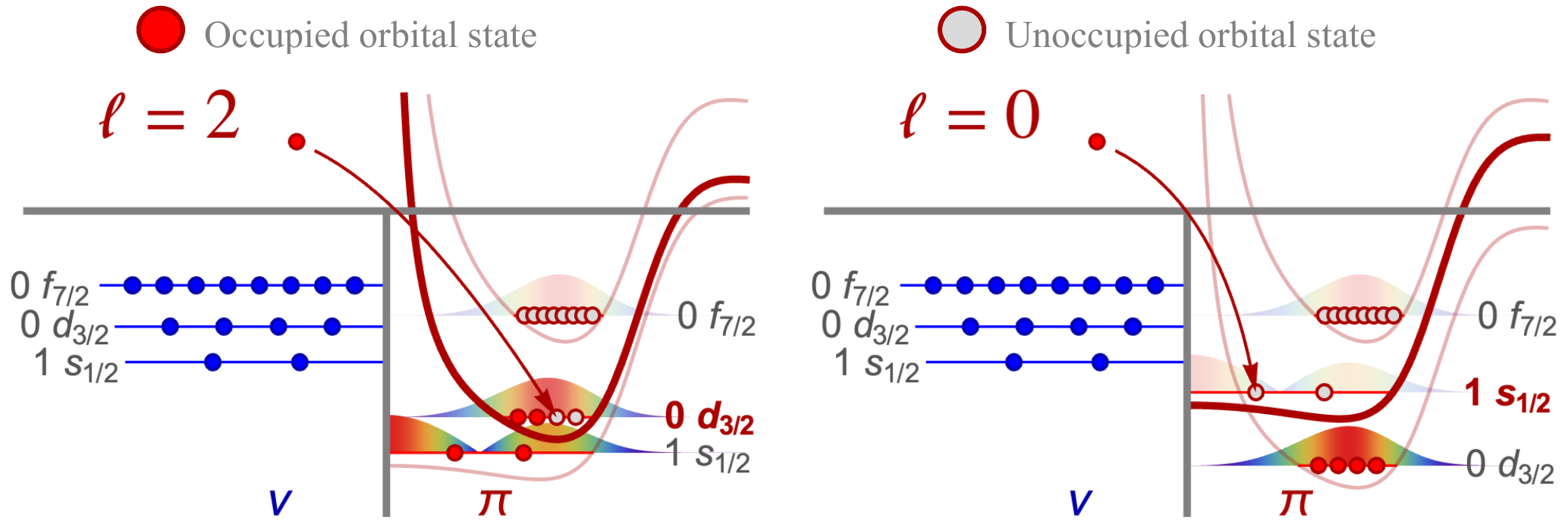}
    \put (5,35) {(a)} \put (55,35) {(b)}
\end{overpic} 
\centering
\hspace{0.2cm}
\begin{overpic}[width=0.18\textwidth]{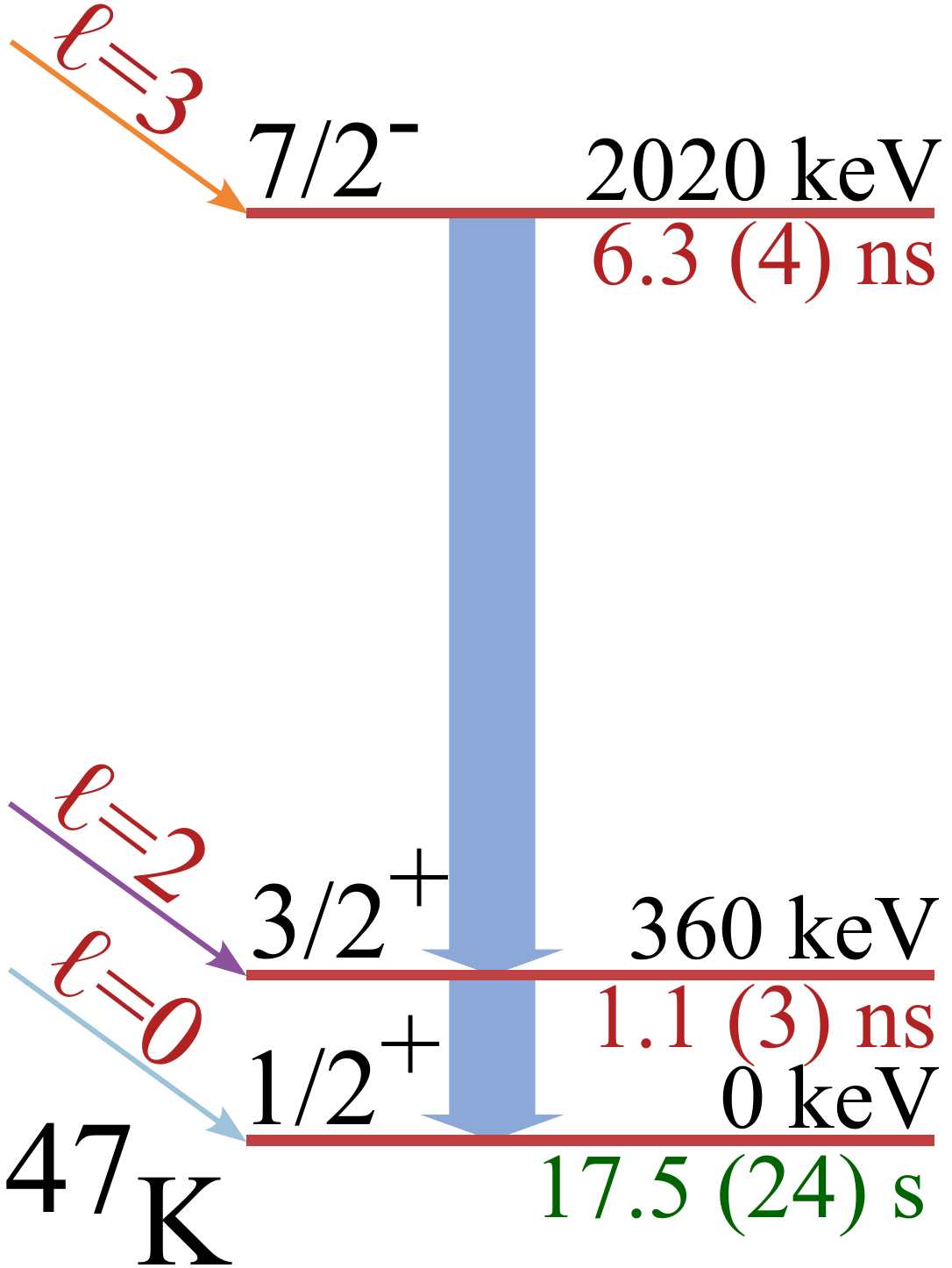}
    \put (18,93) {(c)} 
\end{overpic} 
\caption{Interplay between single particle orbitals, labelled $1s_{1/2}, 0d_{3/2}, 0f_{7/2}$, and their potential wells (red curve) that describe the occurrence of the proton bubble structure. (a,b) Different admixtures of neutron ($\nu$) and proton ($\pi$) single-particle configurations in the wave function of the ground state of $^{46}$Ar affect the proton density distribution. The $1s_{1/2}$ proton orbit favours the localisation of protons in the core of the nucleus due to the absence of the centrifugal term in the potential. A substantial vacancy of protons in this orbital generates the bubble structure in the case of an inverted ordering (b).
(c) Low-lying states of $^{47}$K populated by the $(^3$He$,$d$)$ reaction.}
\label{fig:level_inv_scheme}
\end{figure*}

The charge density distribution of nuclei is a fundamental property on which we can test our understanding of nuclear systems as it is directly related to the saturation of nucleonic matter (that is, to the constraints between the binding energies and radii). 
Charge radius measurements
have recently challenged current theoretical models in exotic isotopes~\cite{Ruiz2016Nat_largeCa,Miller2019Nat_lightCa,Koszorus2021Nat_Kradii}.
Direct measurements on exotic isotopes are still in their infancy~\cite{PhysRevLett.131.092502}, but deviations from this saturated density picture have been conjectured in relation to possible changes in the many-body structure in exotic nuclei. In some cases, information on density profiles can be inferred from other observables. Notably, a recent pioneering study inferred indirect evidence of a charge density depletion, reminiscent of a ``bubble'', situated at the core of the doubly magic $^{34}_{14}$Si isotope, due to a  fully depleted $1s_{1/2}$ orbital~\cite{Mutschler_2016, Duguet2017Si34}.  The $^{46}$Ar isotope, with two proton vacancies in the $^{48}$Ca core, has also been proposed to feature such a charge ``bubble''. However, it would require both an orbital inversion with respect to the standard ordering and the full occupation of the $0d_{3/2}$ (Fig.~\ref{fig:level_inv_scheme}), making this isotope remarkably similar to a closed-shell nucleus. \\
Shell model~\cite{PhysRevC.93.044333,PhysRevC.103.054309,PhysRevLett.77.3967}, single- or multi-reference density functional theory (DFT)~\cite{PhysRevC.80.014317, Wu_2014} have produced contradictory predictions for $^{46}$Ar, , with important discrepancies either among themselves or in comparison with experimental data. 
%
A puzzling open question is the transition probability between the $0^+$ g.s. and the $2^+$ first excited state.
The experimental value of 43(4)~e$^2$fm$^4$~\cite{PhysRevC.93.044333,PhysRevC.68.014302}, is overestimated by a factor greater than three in shell model calculations~\cite{PhysRevC.103.054309,PhysRevLett.77.3967,PhysRevC.93.044333,PhysRevC.68.014302}. 
Some DFT parameterizations predict the standard ordering of energy levels, with a filled $1s_{1/2}$ orbit lying below the partially occupied $0d_{3/2}$ (Fig.~\ref{fig:level_inv_scheme} (a)). Other studies find instead an inversion of the two orbitals that causes nucleons to mostly occupy the $0d_{3/2}$ orbital ($\ell=2$), where the core of the nucleus is strongly interdicted by the centrifugal term of the potential, absent for the for any $s_{1/2}$ orbit. The lack of $\ell=0$ wave protons is the driving mechanism of the central depletion in the core of the nucleus~\cite{Saxena_2019, Shukla_2016, Wu_2014} (Fig.~\ref{fig:level_inv_scheme} (b)). The shell-model Hamiltonian SDPF-U~\cite{PhysRevC.79.014310} predicts an inversion of the orbitals, with almost equal relative occupation of the $0d_{3/2}$, $1s_{1/2}$ orbitals.
%
In this Letter we report on novel proton-addition measurements chosen to directly probe the ground-state structure of $^{46}$Ar, and in particular the occupancies of proton single-particle states. The $1/2^+$ and $3/2^+$ states of $^{47}$K (Fig.~\ref{fig:level_inv_scheme} (c)) represent respectively $s_{1/2}$ and $d_{3/2}$ holes in the closed-shell $^{48}$Ca nucleus, and offer a first hint of the orbital inversion in this isotope. The (${}^3$He,$d$) transfer reaction on $^{46}_{18}$Ar was measured in triple $\gamma$ ray-$d$-$^{47}_{19}$K coincidence and allows for a direct comparison with ab initio calculations, providing firm evidence of a depleted charge density at the centre of this isotope.

\begin{figure}[t]%
\centering
\begin{overpic}[width=0.48\textwidth]{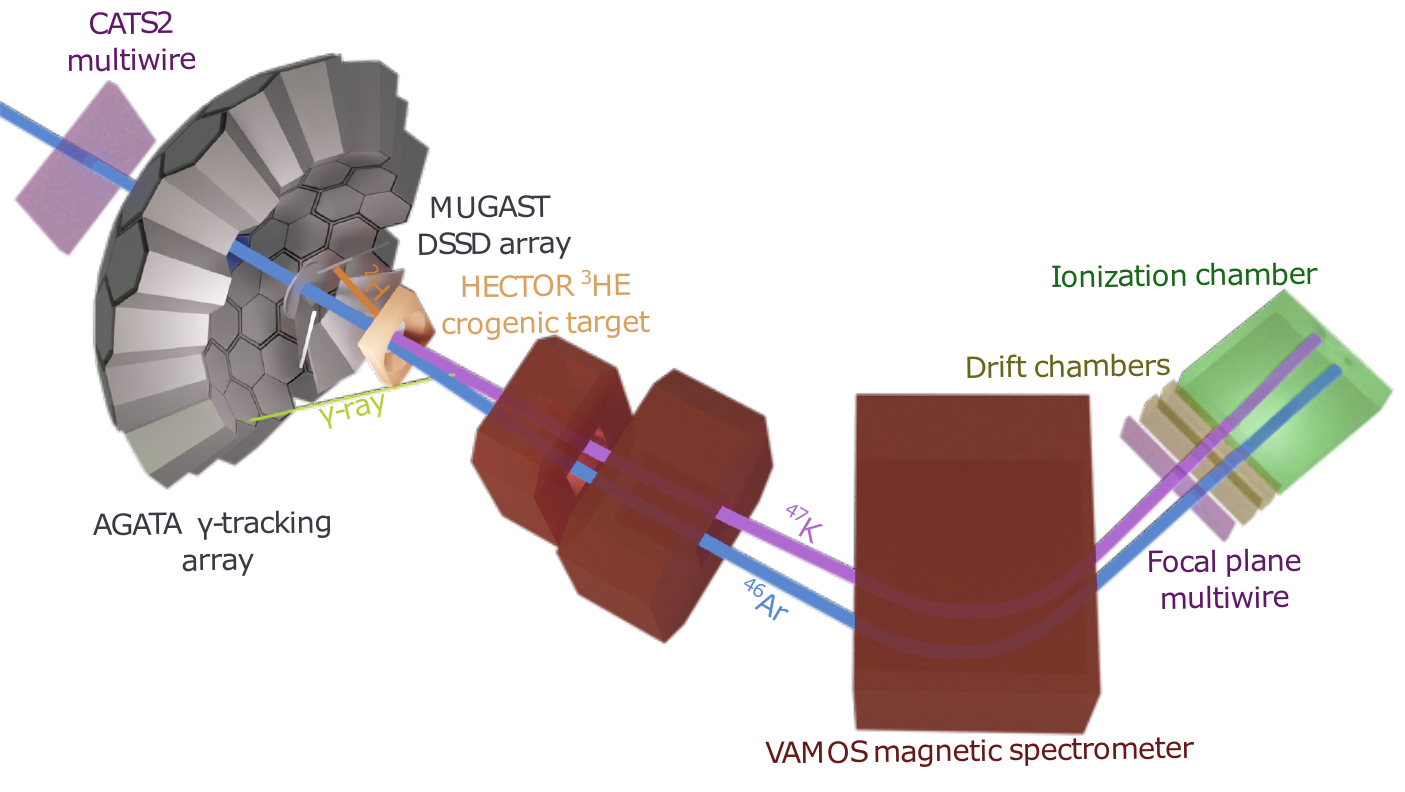}
\end{overpic} 
\caption{
The AGATA-VAMOS-MUGAST experimental setup. Both the deuteron, preferably emitted at backward angles, and the $^{47}$K nucleus, emitted in a small forward cone, were identified. The combination with the $\gamma$ detection allows for the full reconstruction of each event.
}\label{fig:setup_Ex}
\end{figure}

{\itshape Experimental setup.} 
The radioactive, neutron-rich $^{46}$Ar beam was produced at the SPIRAL1 facility in GANIL~\cite{DELAHAYE2020339} by fragmentation of a primary $^{48}$Ca beam impinging on a graphite target at 60 MeV/u. 
The extracted $^{46}$Ar nuclei were post-accelerated to an energy of $9.9$ MeV/u with an intensity of $\approx 5 \times 10^4$~pps and delivered to a cryogenic $^3$He target~\cite{GALTAROSSA2021165830} (Fig.~\ref{fig:setup_Ex}) that achieved an average density of $\approx 1.7$~mg/cm$^2$.

The deuterons, which were predominantly emitted at backward angles as a result of the reaction's kinematics and its differential cross section, were detected using the MUGAST Double-Sided Silicon Strip Detector~\cite{ASSIE2021165743}. 
The array, positioned at laboratory angles ranging from $105^{\circ}$ to $169^{\circ}$ relative to the beam direction, enabled the discrimination of different light particles by correlating their energy with the time-of-flight between the $^{46}$Ar ions passing through the CATS2~\cite{Ottini_Hustache_1999} beam tracker and MUGAST. 
The VAMOS  large-acceptance magnetic spectrometer~\cite{REJMUND2011184} provided mass and atomic number identification of the heavy $^{47}$K recoils on an event-by-event basis. 
A complete characterisation of the reaction products and of the emitted photons was achieved by coupling the setup with the state-of-the-art AGATA $\gamma$-ray tracking array~\cite{AKKOYUN201226,CLEMENT20171}. The tracking capabilities of the array, combined with the complete reaction reconstruction, allowed the high-resolution measurement of the $\gamma$ rays originating from the de-excitation of states in $^{47}$K.

  \begin{figure*}
    \centering
    \begin{tabular}{cc}
    \adjustbox{valign=t}{
          \begin{overpic}[width=0.55\textwidth]{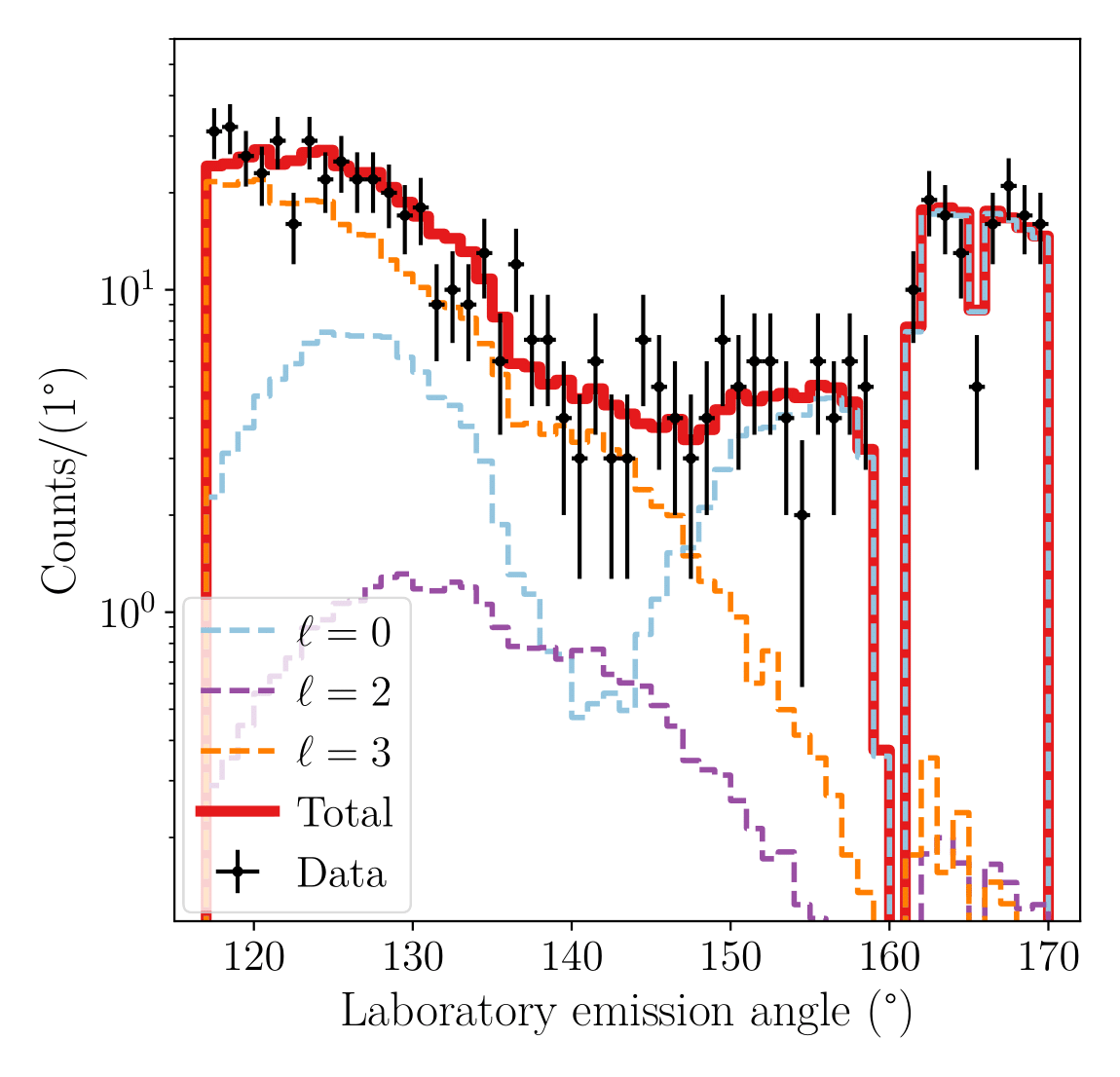}
                \put (90,88) {(a)}
        \end{overpic}
         } 
    &
    \adjustbox{valign=t}{\begin{tabular}{@{}c@{}}
            \begin{overpic}[width=.35\textwidth]{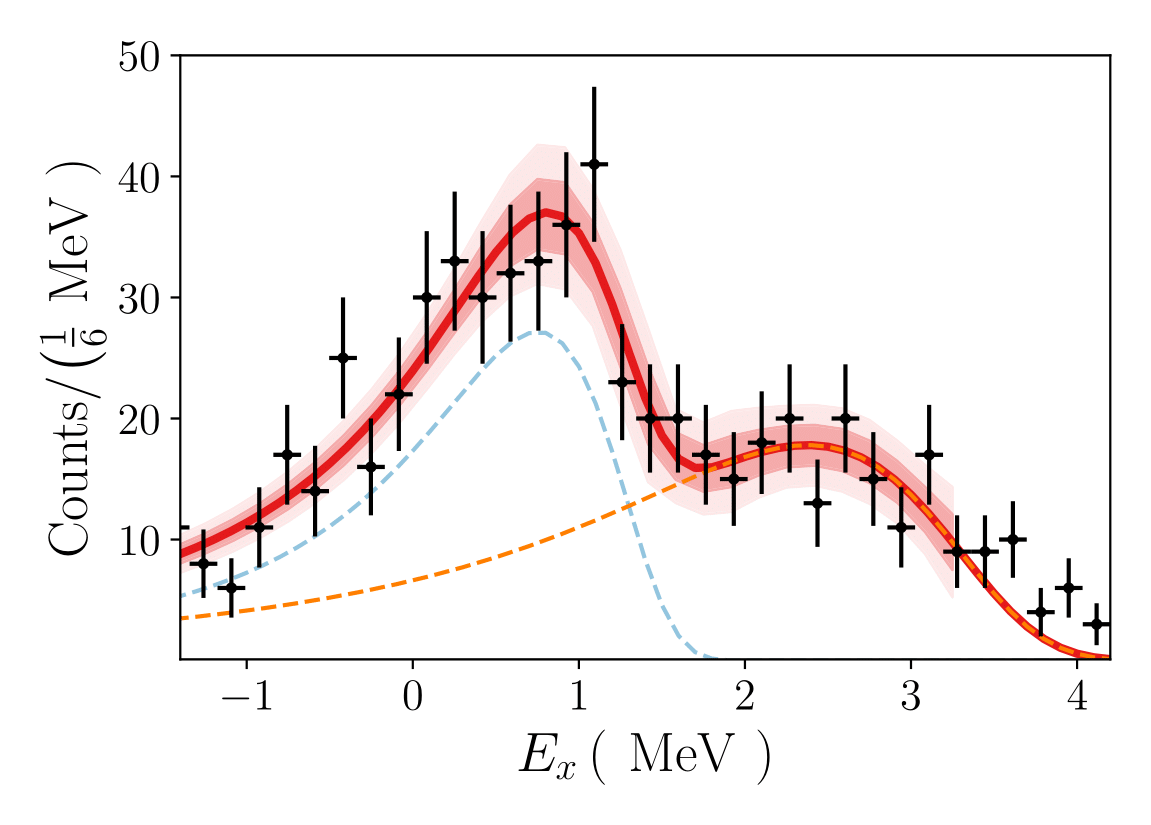}
                \put (86,58) {(b)}
            \end{overpic} \\
            \begin{overpic}[width=.35\textwidth]{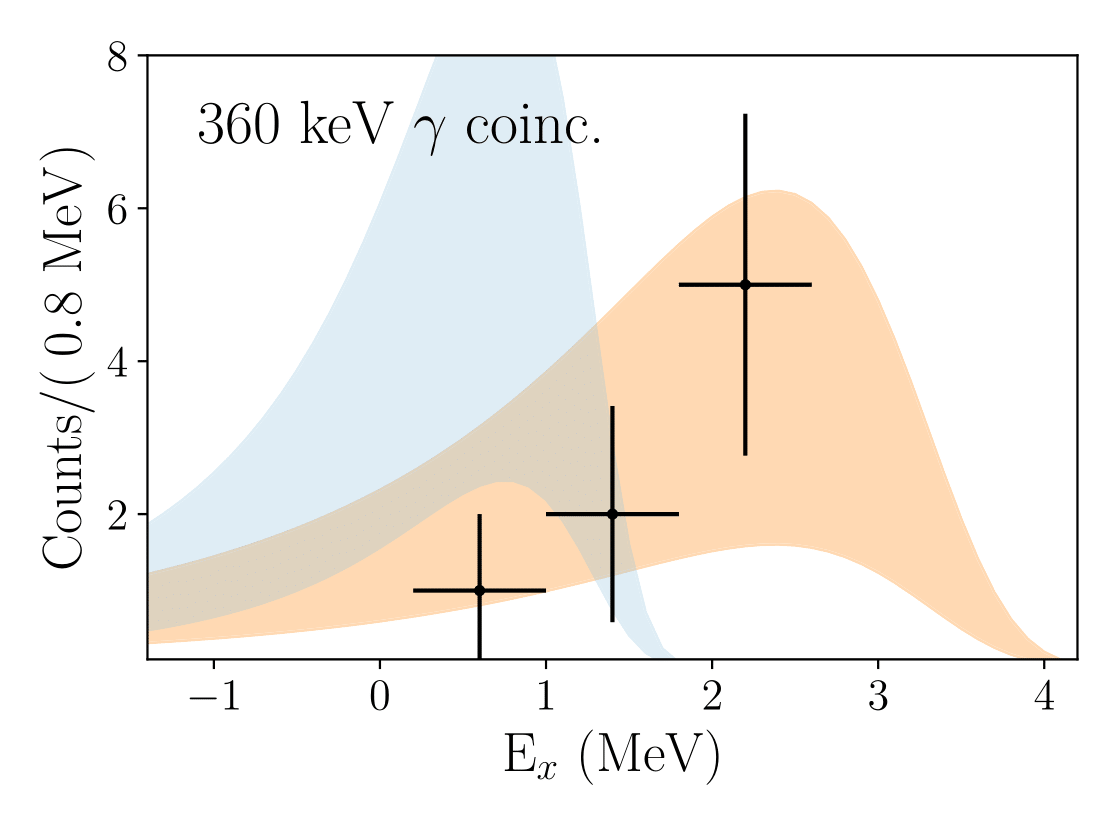}
                \put (86,60) {(c)}
            \end{overpic}
    \end{tabular}} 
    \end{tabular}
    \caption{(a) Histogram of the deuteron angle of emission and result of the maximum likelihood fit (red). The exclusive contribution of different $\ell$-transfer values is shown and indicates a suppressed $\ell=2$ transfer.(b)~Excitation energy spectrum reconstructed from the energy and angle of emission of the deuterons. The fit, in red with 1 and 2-$\sigma$ confidence bands, corresponds to the convolution of two Landau distributions for $\ell=0+2$ (in blue) and $\ell=3$ (orange) wave transfer. (c)~Excitation energy distribution conditioned by the detection of a $360$~keV photon. Shaded areas correspond to the region of interest identified by the distribution without the condition.}\label{fig:fit}
  \end{figure*}

Figure~\ref{fig:fit} (a) shows the angle of emission of the deuterons fitted with the simulated response of the array to different $\ell$-wave transfers ($\chi^2/NDOF\sim0.89$ (C.L.$\sim$80\%)). The response is calculated with a Monte Carlo approach that samples the differential cross section modelled with a finite-range distorted-wave Born approximation (DWBA) for different angular momenta of the captured proton, $\ell$=0, 2 and 3, corresponding to the ground state and first two excited levels in $^{47}$K \cite{SupplMat}.
The parameters of the fitted function are the two spectroscopic factor ratios ${\cal C}^2{\cal S}(\ell=2,3)/{\cal C}^2{\cal S}(\ell=0)$ that scale the single particle cross sections.\\
The resuts, presented in Table~\ref{tab:ratios}, show a suppressed $\ell=2$ wave transfer (in purple in Figure~\ref{fig:fit} (a)). 

\begin{table}[t]
\caption{Ratio of spectroscopic factors ${\cal C}^2{\cal S}(\ell)/{\cal C}^2{\cal S}(\ell=0)$, comparison of experimental and theoretical data.}
\label{tab:ratios}
\begin{ruledtabular}
\begin{tabular}{lccc}
\textrm{Ratio} & \textrm{Exp.} & \textrm{\textit{Ab initio}} & \textrm{SDPF-U~\cite{PhysRevC.79.014310}} \\
\colrule
$\frac{{\cal C}^2{\cal S}(\ell=2)}{{\cal C}^2{\cal S}(\ell=0)}$ & $0.10 \pm 0.10_{\text{stat}} \pm 0.03_{\text{sys}}$ & 0.04 & 0.63 \\
$\frac{{\cal C}^2{\cal S}(\ell=3)}{{\cal C}^2{\cal S}(\ell=0)}$ & $1.10 \pm 0.15_{\text{stat}} \pm 0.17_{\text{sys}}$ & 1.06 
& \\
\end{tabular}
\end{ruledtabular}
\end{table}

Additionally, the excitation energy spectrum (Fig.~\ref{fig:fit} (b)), reconstructed from the angle and energy of the emitted deuteron is fitted ($\chi^2/NDOF\sim 0.79$ (C.L.$\sim$76\%)) with two Lorentzian distributions that correspond to the $(\ell=0)+(\ell=2)$ transfer, in blue, and $\ell=3$ at higher energy. The ratio of the integrals of the two distributions   $Y(\ell=3)/Y(\ell=0)=1.15(42)$, is compatible with the value extracted independently from the angular distributions. The emission of $360$~keV gamma rays can be linked to the population of either of the $7/2^-$ and $3/2^+$ states. These events show an excitation energy only compatible with the direct population of the $7/2^-$ state (\hbox{Fig. \ref{fig:fit} (c)}).
The strongly hindered \hbox{$\ell=2$} transfer cross section represents evidence of a closed $d_{3/2}$ proton subshell in $^{46}$Ar that cannot be filled by adding extra protons.

\begin{figure}[t]%
\centering
\centering
\begin{overpic}[width=0.48\textwidth]{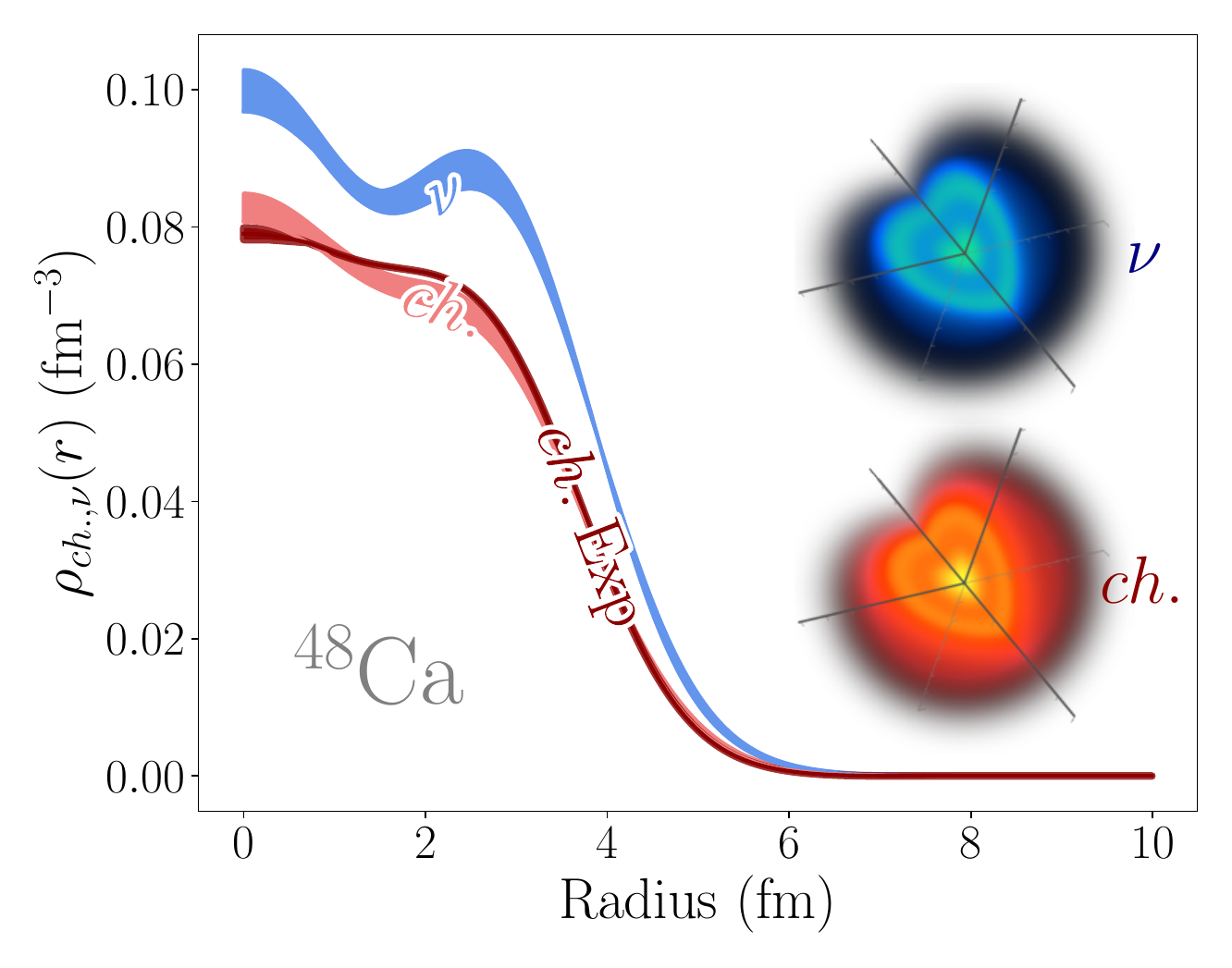}
 \put (19,16) {(a)}
\end{overpic}
\centering
\begin{overpic}[width=0.48\textwidth]{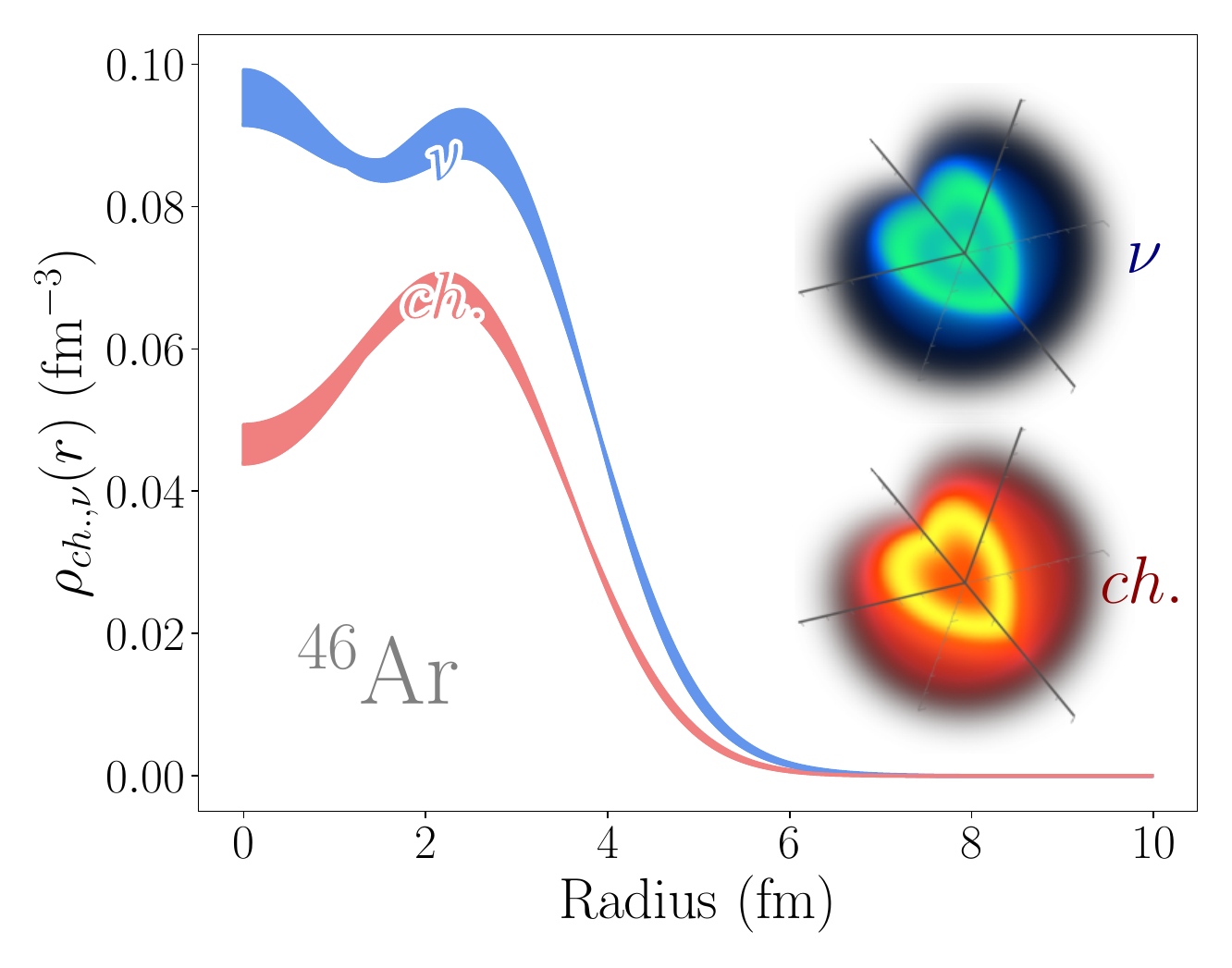}
 \put (19,16) {(b)}
\end{overpic}
\caption{Neutron ($\nu$) and charge density ($ch.$) profiles of $^{48}$Ca (a) and $^{46}$Ar (b). The bands represent the spread of predictions from the four different $\chi$EFT interactions used. The ab initio results are compatible with the measured charge distribution of $^{48}$Ca~\cite{Emrich1983Ca48rho} (red line). For $^{46}$Ar, a charge depletion is predicted and reflects in an empty $s_{1/2}$ orbit at the microscopic level. The three-dimensional density profiles associate brighter colours with higher densities.}
\label{fig:ch_dist}
\end{figure}

{\itshape Ab initio simulations.} To gain further insight into the many-body structure of $^{46}$Ar, we performed self-consistent Green's function (SCGF) computations~\cite{Soma2020Front} using 
a set of four chiral effective field theory ($\chi$EFT) Hamiltonians: the NNLOsat~\cite{Ekstrom2015nnlosat}, the $\Delta$-full $\Delta$NNLO$_{\rm GO}(394)$ and $\Delta$NNLO$_{\rm GO}(450)$~\cite{Ekstrom2018DltGO} and the recently developed 1.8/2.0/(EM7.5)~\cite{Arthuis2024NewMagic}.  These interactions yield highly reliable predictions of known radii and density distributions, as well as reproducing the trend of inversion of $3/2^+$ and $1/2^+$ low-lying states in K isotopes.
The Green’s function was computed in the so-called Algebraic Diagrammatic Construction truncation scheme (ADC($n$)) at order $n$=3~\cite{Barbieri2017LNP,Schirmer2018LNC} that provides consistent information on the one-nucleon addition spectroscopy measured in this work, as well as one-nucleon removal, 
ground-state observables, and the full nucleon density distributions \cite{SupplMat}.

All \emph{ab initio} Hamiltonians confirm that the $1/2^+$ ground state and the $7/2^-$ level of $^{47}$K can be reached directly by proton addition on $^{46}$Ar with large spectroscopic factors ${\cal C}^2{\cal S}$ of 0.64(4) and 0.68(5), respectively (Table~\ref{tab:ratios}). Such values are an indication of empty quasi-particle states in strongly correlated nuclear systems~\cite{Aumann2021PPNP_SFs}. 
Here and in the following, theoretical uncertainties are estimated from model space convergence and variance among all $\chi$EFT Hamiltonians used \cite{SupplMat}.
Our ADC(3) simulations also predict a full $0d_{3/2}$ orbit with spectroscopic factor $\leq$0.03 for proton addition to the second state of $^{47}$K and 0.62(3) for proton removal to a $3/2^+$ state in $^{45}$Cl, further corroborating the interpretation of $^{46}$Ar as a closed shell isotope. 
%
To further validate our simulations, we compare predicted point-neutron and charge density profiles of $^{48}$Ca to the available data in Fig.~\ref{fig:ch_dist} (a). The charge distribution is in good agreement with electron scattering measurements and predicts a root mean square charge radius of 3.505(44)~fm. The predicted neutron skin is 0.148(14)~fm, in agreement with the value of 0.121(35)~fm from measurements of asymmetry in parity-violating electron scattering~\cite{Adhikari2022CREX} and with other \emph{ab initio} simulations~\cite{Hage2016Nat_weakCa}.
Out predicted distributions for $^{46}$Ar are displayed in Fig.~\ref{fig:ch_dist} (b). The theoretical charge radius of 3.462(48)~fm is close to the experiment at 3.4377(44)~fm, while the predicted neutron skin is 0.182(18)~fm. The neutron distribution remains qualitatively unaltered with respect to $^{48}$Ca, while the removal of two protons from the doubly-magic nucleus generates the bubble-like distribution. The depletion factor $F=(\rho_{max}-\rho_{r=0})/\rho_{max}$ highlights a 32(5)~\% drop with respect to the maximum density.
Importantly, the ratios among theoretical spectroscopic factors for proton addition agree with our measurements within $1\sigma$ \cite{SupplMat} and confirm our conclusion of an empty $1s_{1/2}$ orbit,  which in turn is the mechanism responsible for the creation of the charge bubble structure. We have verified that a larger occupation of the proton $1s_{1/2}$, provided it remains within the experimental error, would reduce the internal charge depletion but would  not eliminate it completely (see Supplemental Material~\cite{SupplMat}). 

Mass measurements give further insight on the $0d_{3/2}$ subshell closure for this unconventional structure in $^{46}$Ar.
We consider the 3-point mass difference $\Delta^{(3)} \equiv -\frac{1}2[B(^{47}{\rm K})-2B(^{46}{\rm Ar})+B(^{45}{\rm Cl})]$ where $B(\cdot)$ are binding energies. The value of $2 \Delta^{(3)}$ directly represents the particle-hole proton gap among the empty $1s_{1/2}$ and the fully occupied $0d_{3/2}$ orbits. 
The experimental value $2\Delta^{(3)}_{Exp}=5.55(14)$~MeV~\cite{Wang_2021} for $^{46}$Ar is comparable to the one of doubly-magic $^{48}$Ca and twice the ab initio result of 2.59(28)~MeV, indicating that the proton shell closure in $^{46}$Ar is even stronger than our $\chi$EFT predictions. Note that the $1s_{1/2}$ and $0d_{3/2}$ orbits seen from $^{48}$Ca($e$,$e'p$) reactions almost overlap~\cite{Kramer2001Ca48}, showing a dramatic change of structure with a new proton magic number arising when one moves from Z=20 towards Z=18. The $\chi$EFT prediction of this trend is sound and it is found already at the level of independent particle approximations~\cite{SupplMat}.

We note that standard shell model calculations with the SDPF-U interaction [REF] predict the ground-state configuration as an open shell with a fairly even mixture of the $1s_{1/2}$ and $0d_{3/2}$ proton orbitals but largely overestimate the $^{46}$Ar $B(E2; 0^+ \rightarrow 2^+)$, as seen by the spectroscopic factors in Tab.\ref{tab:ratios}. Considerations based on inelastic proton-scattering (p,p') experiments~\cite{PhysRevC.93.044333} narrowed down this discrepancy to the proton component of the $B(E2)$ matrix element and the presence of too large $1s_{1/2}$ admixtures imposed by the shell model interactions, hence suggesting evidence against an open shell description.
We thus performed new shell model calculations starting with the shell ordering predicted by our ab initio simulations and then mapping the NNLOsat Hamiltonian~\cite{Ekstrom2015nnlosat} into the effective mean-field orbits generated by SCGF-ADC(3), as described in Refs.~\cite{Barbieri2009SFs,Raimondi2019EffCh} and in the Supplemental Material~\cite{SupplMat}.
The obtained small $B(E2)$ value of 35(1)~e$^2$fm$^4$ is in much closer agreement with the experiment and supports a closed shell with a filled proton $0d_{3/2}$ orbit as the solution of the long-standing $B(E2; 0^+ \rightarrow 2^+)$ puzzle.

{\itshape Conclusions.}
The present measurement of the $^{46}$Ar($^3$He,d)$^{47}$K reaction provides evidence of a closed-shell ground state for $^{46}$Ar, with a fully occupied  $0d_{3/2}$ proton orbit and an empty $1s_{1/2}$. The inversion of these two orbits is analogous to that already observed by proton removal from Ca isotopes~\cite{Papuga2014K_inver,Rosenbusch2015K_inver}. However present spectroscopic data, combined with the experimental 3-point mass gap, imply a much larger gap and a new proton magic number at Z=18 around N=28 isotones.

Independent \emph{ab initio} predictions, based on state-of-the-art $\chi$EFT interactions, are in agreement with the experimental findings and provide supporting evidence for a sizeable charge depletion at the center of $^{46}$Ar. To our knowledge, this is the second isotope in which a charge proton bubble is inferred through indirect evidence, after $^{34}$Si~\cite{Mutschler_2016, Duguet2017Si34}.
Our \emph{ab initio} SCGF results confirm the double shell closure of $^{46}$Ar and provide a solution to the puzzle of the small experimental value of the $B(E2; 0^+ \rightarrow 2^+)$.

The presence of a charge bubble is closely connected to the depletion of the $1s_{1/2}$ orbit and its inversion with the $0d_{3/2}$. In this respect, the knowledge of charge density distribution in unstable isotopes, where feasible~\cite{Tsukada2017prl_SCRIT,PhysRevLett.131.092502}, could become a primary tool for discovering regions of unconventional nuclear structure and for advancing our knowledge of nuclear forces.

\vskip .1 in

\emph{Data availability} ---
The data used in this study originate from the E786s GANIL dataset. The ownership of data generated by the AGATA $\gamma$-ray spectrometer resides with the AGATA collaboration as detailed in the AGATA Data Policy~\cite{AGATAdatapolicy}. The source data for the figures is contained in the following reference~\cite{brugnaradataset}.

\vskip .1 in

\emph{Code availability} ---
The code developed for the analysis is contained in reference~\cite{brugnarazenodo}.
All other cited software is available from the authors upon reasonable request.

\vskip .1 in

\emph{Acknowledgments} ---
We acknowledge the GANIL facility for provision of heavy-ion beams and the AGATA, MUGAST and VAMOS collaborations.
This project has received funding from the European Union’s
Horizon 2020 research and innovation programme under grant agreement No. 654002.
We acknowledge support from the PRIN 2017P8KMFT "CTADIR" funding from the Italian Ministry of University and Research.
CloudVeneto~\cite{cloudveneto} is acknowledged for the use of computing and storage facilities.
This research used computational resources at the DiRAC DiAL system at the University of Leicester, UK, (funded  by  the  UK  BEIS via  STFC  Capital  Grants No.~ST/K000373/1 and No.~ST/R002363/1 and STFC DiRAC  Operations  Grant  No.~ST/R001014/1), at the National Energy Research Scientific Computing Center, a DOE Office of Science User Facility supported by the Office of Science of the U.S. Department of Energy under Contract No.~DE-AC02-05CH11231 using NERSC award NP-ERCAP0020946 and of GENCI-TGCC, France, (Contract No. A0130513012).
This work has also been partially supported by the OASIS project no. ANR-17-CE31-0026 and by the U.S. Department of Energy, Office of Science, Office of Nuclear Physics, under contract number DE-AC02-06CH11357, by  MCIN/AEI/10.13039/501100011033, Spain with grant PID2020-118265GB-C42, PRTR-C17.I01, by Generalitat Valenciana, Spain with grant CIPROM/2022/54, ASFAE/2022/031 and by the EU NextGenerationEU, FEDER funds and by MCIN/AEI/10.13039/501100011033, Spain, with grant PID2020-118265GB-C41.

\vskip .1 in

\emph{Author contributions}---
A. G., M. A., F. F., and D. M. designed the experiment.
D. B. performed data analyses, GEANT4 simulations and prepared the figures.
C. B. initially suggested the charge density bubble and led the theoretical simulations.
D. B., C. B., A. G., D. M.  wrote the manuscript.
A. G., M. A., F. F., D. M, A. M., E. C., M. S., A. L.,D. R.,  D. B., F. G., and D. R. prepared the experiment.
A. G.,  S. Bottoni., D. B., G. C., E. V., C. B., V. S., S. Brolli, M. A., F. F.,  contributed to discussion of the interpretation of results and to preliminary draft of the manuscript.
C. B., S. Brolli, V. S. performed \emph{ab initio} SCGF computations.
A. G. and C. B. performed the SM computations.
D. B., S. Bottoni performed the DWBA calculations.
All authors except C. B., G. C., E. V., S. Brolli, V.S. either assisted on the data taking or participated in the development of the experimental apparatus.
All authors discussed the results and commented on the manuscript.

\bibliographystyle{apsrev4-2}
\nocite{thompson2009nuclear, ALLISON2016186, Schiffer1969C2S, PhysRevC.94.054314, osti_4753391, han_pot, BANKS1985381, paxman, Xu2011, Soma2014GkvII, Raimondi2019Resp, Barbieri2022GADC3, Horowitz2012rSO, Cipollone2015prc, Angeli2013Rch, Burrows2007NDSK47, Lapoux2016prl, Arthuis2020Xe, Soma2020LNL, Barbieri2019nuAr, Bai2010Sly5Tw5} 
\bibliography{sn-bibliography}

\pagebreak
\widetext
\begin{center}
\textbf{\large Supplemental Materials}
\end{center}

\subsection*{Data Analysis and Results}

The analysis strategy exploits the strong dependence of the differential cross section on the angular momentum $\ell$ of the transferred nucleon in direct reactions.  In particular, the addition of a proton to an $s$-wave ($\ell=0$) single-particle orbital leads to a distribution of the ejected deuteron that is peaked at backward angles in the laboratory frame of reference. This feature generates the signature that allows for a clear comparison with the distribution of $f$ ($\ell=3$) and $d$ ($\ell=2$) wave transfers.

The deuterons angular distributions of the direct reactions were computed in the finite-range DWBA approximation with the FRESCO code~\cite{thompson2009nuclear}.

\begin{figure}[th]%
\includegraphics[width=0.5\textwidth]{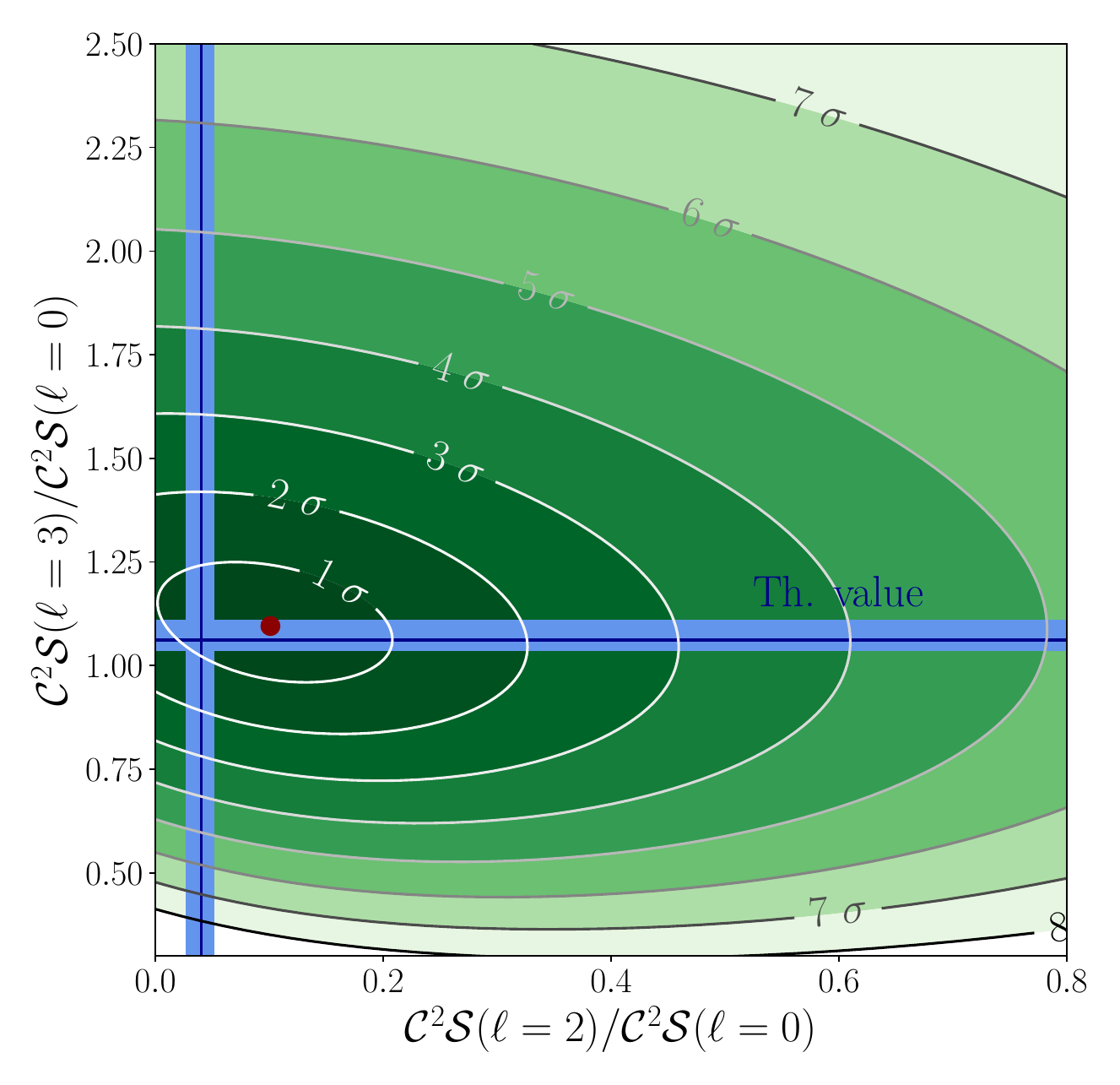}
\caption{Comparison of theoretical and experimental spectroscopic factors. The horizontal (vertical) axes correlate the amount of $\ell=2$ ($\ell=3$) over the amount of transfer to the ground state, $\ell=0$. The theoretical uncertainty bands consist in the interval between the minimum and maximum predicted value for the ratio of spectroscopic factors with the different parametrizations discussed in the main body (see Table~\ref{tab:th_sf}).}\label{fig:lh}
\end{figure}

The differential cross sections serve as an input for a Monte Carlo GEANT4~\cite{ALLISON2016186} simulation that extracts the response of the experimental apparatus for the population of the three different states of $^{47}$K: $\ell=0$ transfer to the $1/2^+$ g.s., $\ell=2$ transfer to the $3/2^+$ state and $\ell=3$ to the $7/2^-$ ( Fig. 1 (c) of the main article). The spectroscopic factors are extracted with a maximum likelihood fit considering the following relationship with the inclusive differential cross section~\cite{Aumann2021PPNP_SFs}, where $k$ indexes the populated states:
\begin{equation}
 \frac{d \sigma}{d \Omega} = \sum_k g_k\; \mathcal{C}^2\mathcal{S}_k \; \frac{d \sigma^{SP}_k}{d \Omega} \; ,
    \label{eq:single_part_cross_sec}
\end{equation}
Here $\sigma^{SP}_k$ is the theoretical cross section relative to a single-particle orbit, $g$ represents the statistical factor and equals the orbital degeneracy, $(2j+1)$, for particle addition and $\cal C$ is the product of an appropriate angular Clebsh-Gordan coefficient and an analogous isospin term~\cite{Schiffer1969C2S,PhysRevC.94.054314}. In the case of the $^{46}$Ar($^3$He,d)$^{47}$K reaction $\mathcal{C}^2 \approx 1$. The product $\mathcal{C}^2\mathcal{S}_k$ is the spectroscopic factor and it enters in Eq.~\ref{eq:single_part_cross_sec} as
as a modulation factor that can be interpreted within a theoretical framework as a fraction of the full orbital occupation.  
The maximisation of the likelihood is performed on the experimental distribution of the emission angle of the deuteron while the excitation energy serves as an independent evaluation. 

The likelihood profile for the experimental ratios ${\cal C}^2{\cal S}(\ell=2)/{\cal C}^2{\cal S}(\ell=0)$  and  ${\cal C}^2{\cal S}(\ell=3)/{\cal C}^2{\cal S}(\ell=0)$ is displayed in Fig.~\ref{fig:lh} along with the results of SCGF ab initio simulations discussed in the main text and in the next sections, also presented in Table~\ref{tab:th_sf}.

\begin{table}[]
\begin{tabular}{lll}
                   & $\frac{{\cal C}^2{\cal S}(\ell=2)}{{\cal C}^2{\cal S}(\ell=0)}$    & $\frac{{\cal C}^2{\cal S}(\ell=3)}{{\cal C}^2{\cal S}(\ell=0)}$    \\
\textbf{NNLOsat}            & 0.028    & 1.049 \\
\textbf{$\Delta$NNLO${}_{\text{GO}}$(394)} & 0.033    & 1.109    \\
\textbf{$\Delta$NNLO${}_{\text{GO}}$(450)} & 0.049    & 1.059    \\
\textbf{1.8/2.0(EM7.5)}    & 0.051    & 1.036    \\
Average            & 0.040    & 1.063   
\end{tabular}
\caption{Theoretical results with different parametrizations of the interaction}
\label{tab:th_sf}
\end{table}

\subsection*{Optical Potentials}

The global optical potentials by Becchetti et al. \cite{osti_4753391} and by Han et al. \cite{han_pot} were adopted for the entrance channel $^3$He-$^{46}$Ar and for the exit channel d-$^{47}$K, respectively. 
These potentials provide, among those present in literature, respectively the best fit for all the transfer channels of the mirror reaction $^{48}$Ca(d,$^3$He)$^{47}$K~\cite{BANKS1985381} and for the elastic deuteron scattering on $^{47}$K~\cite{paxman}. In particular, they were found to account correctly for the attenuation of the cross section at higher angles in the centre of mass of the angular range of sensitivity.

Some experimental uncertainties affect in a coherent manner different $\ell$-wave transfers and are eliminated or suppressed when only relative values of spectroscopic factors are considered. The statistical treatment of the results is self consistent as it only involves ratios of cross sections and spectroscopic factors.
As an example, the uncertainties introduced by the gas density at temperatures close to the critical point of $^3$He and the integrated beam current are elided. Additionally, the ratio has the advantage of also minimizing the dependence on the optical potential choice. Although the choice of the parametrization is based on experimental data, the evaluation of the systematic dependence on the parametrization has been evaluated by performing the statistical analysis on the second best fitting $^3$He optical potential~\cite{Xu2011}.

The outgoing $d$ potential has negligible effects on the outcome.

\subsection*{$\gamma$-ray Coincidence Analysis}

The detection of discrete $\gamma$-rays offers the possibility for an independent analysis with respect to the angular distribution and the excitation energy. Neglecting the condition of detecting deuterons with the silicon detector, the reactions contributing to the observed gamma rays include not only the $(^3$He,d$)$, but also the $(^3$He,pn$)$ channel.

A Monte Carlo GEANT4 simulation has been developed to evaluate the transparency of the copper target cell to $\gamma$-rays and the dynamics of the reaction~\cite{GALTAROSSA2021165830}. The simulation has been validated with data taken with a $^{152}$Eu source placed $85$~mm downstream to mimic the emission of photons from long-lived states.

\begin{figure}[t]%
\centering
\includegraphics[width=0.5\textwidth]{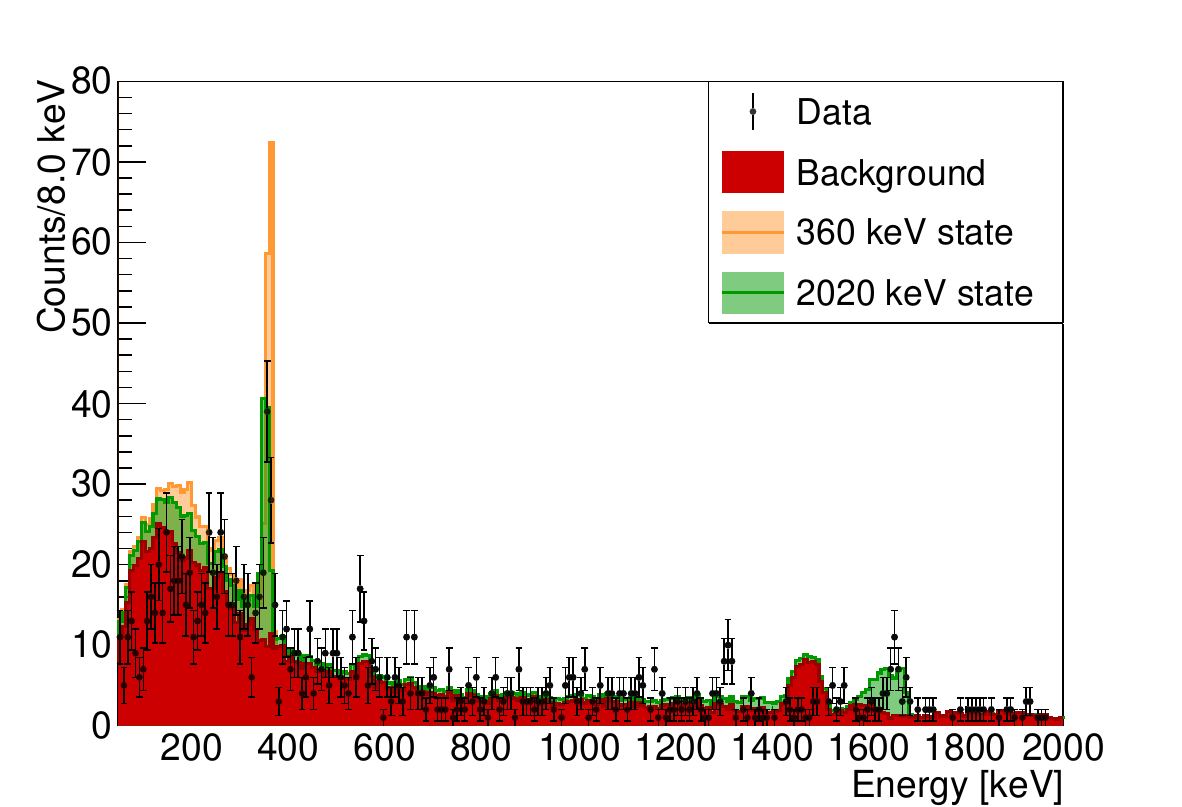}
\caption{Doppler-corrected $\gamma$-ray spectrum in coincidence with $^{47}$K detected by the magnetic spectrometer. The data is compared with a simulation of the pure population of the $3/2^+$ (in yellow) and $7/2^-$ state (in green). The data is compatible with a negligible population of the first excited state.}\label{fig:agata-vamos-gamma}
\end{figure}

The probability of detecting $360$ keV gammas rays emitted from the de-excitation of the $3/2^+$ state differs significantly if the state is populated directly. This effect is caused by the long half-life of the $7/2^-$ state ($6.3(4)$ ns) that branches to the $3/2^+$ when compared to the significantly lower value of $1.1(3)$ ns of the $3/2^+$ state (Fig. 1 of the article). This effect, combined with the velocity of the recoil $\beta \approx 13\%$, changes significantly the mean angular coverage of the detector, impacting its absolute efficiency.
Fig.~\ref{fig:agata-vamos-gamma} shows the response of the detector in case of direct transfer to the $3/2^+$ state (in yellow) and the $7/2^-$ (in green). The experimental data indicates that the detected $360$-keV photons are compatible with the population of higher-lying states. 
The accumulation of counts present around $1500$ keV corresponds to the emission of a $1345(3)$ keV photon by isotopes of $^{46}$K implanted in the chamber. The remaining deviation from the simulated response peak, at $1319$~keV, is consistent with the population of a $^{47}$K state at $3350(30)$~keV ($5/2^+,3/2^+$), which decays to the $7/2^-$ state.

The observed response leads to the ratio of spectroscopic factors equal to $\mathcal{C}^2\mathcal{S}[\ell=2]/\mathcal{C}^2\mathcal{S}[\ell=3] =0.10 \pm 0.36$, representing an independent confirmation of the value extracted from the angular distribution.

\subsection*{SCGF Calculations}

The nuclear quantum many-body problem is solved with the self-consistent Green's function (SCGF) method, see Refs.~\cite{Barbieri2017LNP,Soma2014GkvII,Soma2020Front} for computational details and an overview of applications to nuclear physics. The  one-body propagator, or Green's function, is obtained as a solution of the Dyson equation
\begin{equation}
   g_{\alpha\beta}(\omega) =  g^{(OpRS)}_{\alpha\beta}(\omega) 
   + \sum_{\gamma\,\delta}     g^{(OpRS)}_{\alpha\gamma}(\omega)  \Sigma^{\star}_{\gamma\delta}(\omega)  g_{\delta\beta}(\omega),
\end{equation}
where greek indices $\{\alpha,\beta,\ldots\}$ label single-particle states of the model space and $g_{\alpha\beta}^{(OpRS)}(\omega)$ is an unperturbed propagator, which we dub as {\it optimised reference state} (OpRS). 
The latter is generated self-consistently from the spectral function associated with the true solution $g_{\alpha\beta}(\omega)$ by imposing that it reproduces the zeroth and first moments with respect to the inverse frequency, $[\omega-E_F]^{-1}$, where $E_F$ is the Fermi energy~\cite{Raimondi2019Resp,Barbieri2022GADC3}. In practice, one starts from an approximate mean-field propagator, for example, a Hartree-Fock, and updates the OpRS at each iteration: the converged $g_{\alpha\beta}^{(OpRS)}(\omega)$ is completely independent of the choice of the starting point~\cite{Soma2014GkvII}.

The knowledge of $g_{\alpha\beta}(\omega)$ allows to compute
the ground-state energy $E_0^A$ (using the Koltun sum rule) and any one-body observable associated with the A-nucleon isotope, including the density matrix and point-nucleon radii~\cite{Barbieri2017LNP}. The $g_{\alpha\beta}(\omega)$ also incorporates all the information on the spectroscopy of neighbouring $^{47}$K and $^{45}$Cl isotopes presented in the main text.

Many-particle correlations are encoded in the self-energy $\Sigma_{\alpha\beta}^\star(\omega)$. We compute it using the algebraic diagrammatic construction method, or ADC($n$), which provides a systematically improvable hierarchy of many-body truncations~\cite{Barbieri2017LNP,Schirmer2018LNC}. At level $n=2$ it resums second-order perturbation theory contributions, while $n$=3 includes complete all-order resummations of ring diagrams and particle-particle, hole-hole and particle-hole ladders. ADC(2) computations are already sufficient for converging nucleon density distributions while ADC(3) typically predicts ground-state energies within a few \% and it allows for reliable predictions of nucleon addition and removal spectroscopy.

The root mean square (rms) point-proton and point-neutron radii are computed from the intrinsic operators $R_p^2\equiv\frac 1 Z \sum_{p=1}^Z (\vec{r}_p-\vec{R}_{cm})^2$ and $R_n^2\equiv\frac 1 N \sum_{n=1}^N (\vec{r}_n-\vec{R}_{cm})^2$. Here, the index $p$ runs over all the Z protons, the index $n$ runs over the N neutrons, A=N+Z is the total number of nucleons and $\vec{R}_{cm} = \frac 1 A \sum_{i=1}^A \vec{r}_i$ is the centre of mass. 
In our ab initio calculations we first compute the point-nucleon rms radii and then obtain the neutron skin as $R_p - R_n$ and the charge radius as $R_{ch}^2 = R^2_p + \langle r_p^2 \rangle+(N/Z) \langle r_n^2\rangle+(3/4m_N^2) + \langle r^2\rangle_{so}$. Here  $\langle r_p^2\rangle$ = 0.7079 fm$^2$ is the mean squared charge radius of the proton, $\langle r_n^2\rangle$= -0.1149 fm$^2$ is that of the neutron, ($3/4m_N^2$) = 0.033 fm$^2$ is the relativistic Darwin–Foldy correction, and $\langle r^2\rangle_{so}$ is the spin–orbit correction~\cite{Horowitz2012rSO}. The charge density distribution is computed by convoluting the point-proton density distributions with the proton and neutron charge densities, as explained in Ref.~\cite{Duguet2017Si34}.

\subsection*{Choice of the Hamiltonian and Model Space Dependence}

The SCGF computations employ the intrinsic Hamiltonian with two-nucleon (2N) and three-nucleon (3N) interactions,
\begin{align}
  H = \sum_i  \frac{p_i^2}{2 m_N} - T_{C.M.}^{[A]} \!+\! \sum_{i<j}  V^{(2N)}_{i,j} \!+\!\! \sum_{i<j<k}  W^{(3N)}_{i,j,k}  ,
\nonumber
\end{align}
where $\vec{p}_i$ is the momentum of nucleon $i$, $m_N$ is the nucleon mass and $T_{C.M.}^{[\cdot]}$ is the operator for the kinetic energy of the centre of mass that depends on the number of particles. We evaluate it for A=46 nucleons when computing ground-state properties of $^{46}$Ar and for A=47 and 45 to compute nucleon addition to $^{47}$K and removal to  $^{45}$Cl, respectively~\cite{Cipollone2015prc}. 
All operators are expanded on a spherical harmonic oscillator (HO) basis containing 
up to 14 shells (or $N_{\rm max}$ = 13), 
keeping all matrix elements of one- and two-body operators and limiting the 3N interactions to configurations characterized by $N_1 +N_2+N_3  \leq E_{\rm 3max} =$ 16. 

\begin{table*}[]
\centering
\renewcommand{\arraystretch}{1.3} 
\begin{tabular}{l|c|c|c|c|c|c}
\hline
& \textbf{NNLOsat} & \textbf{$\Delta$NNLO${}_{\mathrm{GO}}$} & \textbf{$\Delta$NNLO${}_{\mathrm{GO}}$} & \textbf{1.8/2.0} &  Average & Experiment \\ 
&          &  \textbf{($\Lambda$=394)}  & \textbf{($\Lambda$=450)}   & \textbf{(EM7.5)} &   \\ \hline
{\bf ${}^{48}\text{Ca}$:} &    &       &        &        &   \\
$r_{ch}$ (fm) & 3.494(46) & 3.495(15) & 3.480(36) & 3.551(5) & 3.505(44) & 3.4771(20)\,\cite{Angeli2013Rch}  \\ 
$\nu$-skin (fm) &   0.135(10)  &  0.148(11)  & 0.149(10)   &  0.162(3)  &  0.148(14) & 0.121(35)\,\cite{Adhikari2022CREX} \\ 
$\Delta E^{(h)}_{{3/2}^{+} \!-\! {1/2}^{+}}\!$ (MeV)\! &   1.41(27)    &  1.00(8)    &  1.21(21)  &  1.70(56)  &  1.33(45) & 0.360~\cite{Burrows2007NDSK47} \\
B(${}^{48}$Ca) (MeV) & 415.0   &  418.7 &  427.9  &  414.6   & 419.0(6.2) & 416.00\,\cite{Wang_2021} \\ 
{\bf ${}^{46}\text{Ar}$:} &    &       &        &        &   \\
$r_{ch}$ (fm) & 3.443(43)  & 3.453(2)   & 3.434(39) & 3.519(4)  &  3.462(48) & 3.4377(44)\,\cite{Angeli2013Rch}  \\ 
$\nu$-skin (fm) &  0.169(16)  & 0.179(18) &  0.188(13) &  0.194(5) & 0.182(18)  \\ 
F (\%) & 37(5)   &   31(3)    &    30(4)    &   28(2)     & 32(5)   \\ 
$2 \Delta^{(3)}$ (MeV) & 2.67(28) & 2.34(7) & 2.52(23) & 2.81(8) & 2.59(28) & 5.56(14)\,\cite{Wang_2021}   \\ 
B(${}^{46}$Ar) (MeV) &  384.2   &  386  &  393.6   &  381.5  & 386.3(5.2) & 386.97\,\cite{Wang_2021}  \\
{\bf ${}^{46}$Ar $\to {}^{47}$K:}  &   &       &        &        &   \\ 
    ${\cal C}^2{\cal S}_{1/2^{+}(g.s.)}$   &  0.68(2)  &  0.63(2)  &  0.64(3)  &  0.60(1)   & 0.64(4)  \\
${\cal C}^2{\cal S}_{3/2^{+}}$  & 0.019(6) & 0.021(5)  &  0.032(6) &  0.031(1) &  0.026(8) \\ 
${\cal C}^2{\cal S}_{7/2^{-}}$   & 0.71(2)  & 0.070(2) & 0.68(3) & 0.63(2) & 0.68(5) \\ 
{\bf ${}^{46}$Ar $\to {}^{45}$Cl:} &           &       &        &        &   \\  
${\cal C}^2{\cal S}_{3/2^{+}(g.s.)}$   &  0.65(2)  &  0.63(2)  & 0.61(3)  &  0.58(1)   & 0.62(3)  \\
  \hline
\end{tabular}
%
\caption{Breakdown of SCGF predictions from the complete Dyson-ADC(3) propagator, $g_{\alpha\beta}(\omega)$, and for each of the four $\chi$EFT Hamiltonian employed. Values are displayed for charge radii ($r_{ch}$), neutron skins ($\nu$-skin),
the three-point gap formulas ($\Delta^{(3)}$) for $^{46}$Ar, binding energies (\,B($\cdot$)\,) and the depletion factor ($F$) for the proton charge bubble of ${}^{46}$Ar.
The $\Delta E^{(h)}_{{3/2}^{+} \!-\! {1/2}^{+}}\!$ is the energy gap among dominant quasihole peaks for proton removal from ${}^{48}$Ca.
For one-proton addition~(removal) to ${}^{47}$K~(${}^{45}$Cl) we provide 
the spectroscopic factors (${\cal C}^2{\cal S}$) expressed as a fraction of the full orbit occupation, $2j+1$.  For each interaction, the errors are estimated from model space uncertainties. The errors on the average theoretical predictions include the standard deviation among all Hamiltonians (see text). The known experimental values of observable quantities are given in the last column.
}
\label{Tab:XEFTvalues}
\end{table*}

The present analysis is based on four distinct chiral effective field theory ($\chi$EFT) Hamiltonians. We performed computations with the next-to-next-to-leading order (NNLO) interactions denoted as NNLOsat~\cite{Ekstrom2015nnlosat},
with two instances of the $\Delta$-full NNLO interactions from Ref.~\cite{Ekstrom2018DltGO} with different momentum cutoffs that are denoted as $\Delta$NNLO$_{\rm GO}$(394) and $\Delta$NNLO$_{\rm GO}$(450) and with a N3LO two-body interaction complemented with NNLO three-body operators proposed in Ref.~\cite{Arthuis2024NewMagic} that is denoted 1.8/2.0/(EM7.5).
While based on different degrees of freedom, resolution scales and fitting procedures, all these forces stand out among the chiral interactions because they lead to a satisfactory simultaneous reproduction of nuclear radii and binding energies. This is key in the current study, which focuses on nuclear densities.
The NNLOsat, in particular, has been found to reproduce very accurately the known density distributions for various isotopes, including $^{16}$O~\cite{Lapoux2016prl},  $^{36}$S~\cite{Duguet2017Si34},  $^{40,48}$Ca~\cite{Hage2016Nat_weakCa} and $^{132}$Xe~\cite{Arthuis2020Xe}. The experimental trend of inversion and re-inversion of the low-lying $3/2^+$ and $1/2^+$ states in $^A$K isotopes with varying neutron number is reproduced by NNLOsat~\cite{Soma2020LNL} and we have verified that the same holds for the other Hamiltonians.
For the central values reported below and in the main text we follow Refs.~\cite{Barbieri2019nuAr,Hage2016Nat_weakCa} and adopt $\hbar\Omega$=20 MeV for the NNLOsat and the $\Delta$NNLO$_{\rm GO}$ interactions as the oscillator frequency that best reproduces the known charge distribution for neighbouring nuclei such as $^{36}$S~\cite{Duguet2017Si34} and $^{48}$Ca~\cite{Hage2016Nat_weakCa}. Only for 1.8/2.0/(EM7.5), we used $\hbar\Omega$=16 MeV~\cite{Arthuis2024NewMagic}.
We estimate the theoretical errors associated with model space dependence and with both the accuracy and precision of the $\chi$EFT Hamiltonians as follows. First, we follow Ref.~\cite{Arthuis2020Xe} and perform additional SCGF computations with varying model space sizes, $N_{\rm max}$, and frequencies, $\hbar\Omega$, to find the combination for the best convergence of radii. The difference with the adopted central values, the latter being closer to the experiment by construction, are taken as a combined estimate of the uncertainties associated with model space convergence (which is smaller) and the accuracy of the interaction.
Note that model-space converged radii differ by no more than 1-2\% from the experiment for the soft and saturating nuclear forces employed here (see also Refs.~\cite{Soma2020LNL,Koszorus2021Nat_Kradii} and errors in Table~\ref{Tab:XEFTvalues}), which represents the typical accuracy in state-of-the-art ab initio theory.
Next, we compute the average and standard deviation among the different nuclear forces to estimate the precision from $\chi$EFT. The final error is obtained by summing all uncertainties in quadrature. Because of our choice of using the $\hbar\Omega$ that is in better agreement with the experimental radii in $^{48}$Ca, our final theoretical error can be considered as being a conservative one. 
Table~\ref{Tab:XEFTvalues} displays the breakdown of our predictions from each nuclear interaction for the various observables discussed in the text. 

All simulations predict $^{46}$Ar as a $0d_{3/2}$ closed shell with an empty $	1s_{1/2}$.  Likewise, variations of $\hbar\Omega$ and $N_{\rm max}$ do not appreciably alter the presence of a charge density bubble. All theoretical errors indicated in the main text result from the above analysis that combines the $\hbar\Omega$ dependence, model spaces with 12 and 14 oscillator shells and the comparison among different $\chi$EFT Hamiltonians.

\subsection*{Shell Structure of $^{46}$Ar and $^{48}$Ca}

\begin{figure}[h!]%
\includegraphics[width=.5\textwidth]{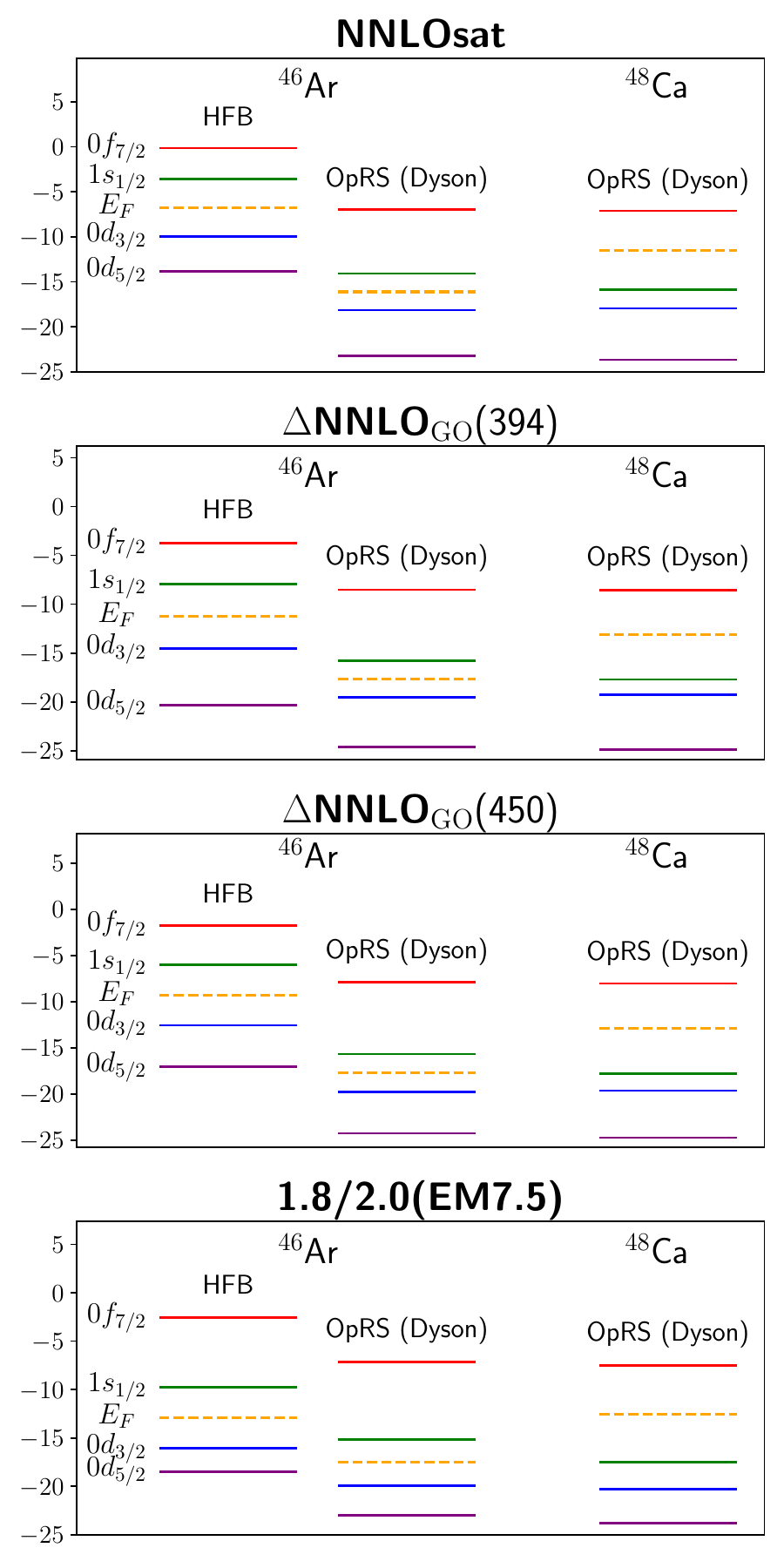}
\caption{Proton energies associated to various independent-particle approximations of ${}^{46}$Ar and ${}^{48}$Ca. All spectra are computed with the four saturating chiral Hamiltonians presented in the main text. The HFB are mean-field computations that allow for pairing effects. The OpRS includes the averaged effects of full many-body correlations and are deduced from the spectral functions computed with SCGF and the Dyson-ADC(3) approach.}\label{fig:EnLev}
\end{figure}
The splitting among $1s_{1/2}$ and $0d_{3/2}$ in $^{48}$Ca can be estimated by the difference among the separation energies to dominant quasihole states, $\Delta E^{(h)}_{{3/2}^{+} \!-\! {1/2}^{+}}\!$. By contrast, the same splitting for $^{46}$Ar is experimentally reflected in the 3-point mass formula and precisely by $2 \Delta^{(3)}_{^{46}\rm Ar}$. The full Dyson-ADC(3) predictions for these quantities are compared to experiment in Table~\ref{Tab:XEFTvalues}.
Further insight into the proton shell structure of the two isotopes can be gained from various independent particle approximations of their wavefunctions. These are shown in Figure~\ref{fig:EnLev}, separately for each $\chi$EFT interaction.
The left columns are Hartree-Fock-Bogoliubov (HFB) mean-field results. Computations with the HFB and with the Gorkov-ADC(2) approximation (not shown here) are both capable to account for pairing effects. However, they predict the absence of pairing in this particular case as well as a sizeable gap between $1s_{1/2}$ and $0d_{3/2}$, the latter being more bound. 
The OpRS spectra is very close to the centroids of the fragmented spectral distributions. This is a much better representation of the shell structure than the Hartree-Fock(-Bogoliubov) mean field since it includes averaged effect full many-body correlation that lead to additional binding and compression of the spectra.
All simulations and all interactions display a clear trend with the inversion of $1s_{1/2}$ and $0d_{3/2}$ at N=28 and their gap becoming larger when removing two protons from $^{48}$Ca to $^{46}$Ar.

\subsection*{Shell Model Calculations}

The OpRS resulting from ab initio simulations are constructed to reproduce very closely both the exact nucleon density distributions and the total ground-state energy (through the Koltun and Koopman's sum rules) predicted by $g_{\alpha\beta}(\omega)$~\cite{Barbieri2022GADC3}. Hence, they are deemed to be an improved approximation for a reference mean field and an ideal representation of the shell structure on which residual valence-space interactions can be built.
To perform shell model calculations based on $\chi$EFT Hamiltonians, we select from $g_{\alpha\beta}^{(OpRS)}(\omega)$ the OpRS orbits and single particle energies corresponding to the $1s\,0d\,0f\,1p$ model space. The NNLOsat interaction is then mapped into a new effective shell model interaction by computing its matrix elements with respect to the new basis, as discussed in Refs.~\cite{Barbieri2009SFs,Raimondi2019EffCh}.
The clear advantage of this approach is that the effective interaction and charges employed in the shell model computations are derived directly from the $\chi$EFT interaction and by extension from the underlying symmetries of QCD. Hence, it removes the main source of phenomenology while providing a shell model calculation that is consistent with the ab initio Hamiltonians discussed above.

The B(E2) transitions were computed with the ANTOINE shell model code~\cite{Caurier2005rmpSM}, leaving the $1s\,0d$ fully open and allowing up to 5 nucleon excitations either across the proton $1s\,0d$ and $1p0f$ shells or above the neutron $0f_{7/2}$ orbit. Computations were based on the OpRS model space and interactions extracted with $\hbar\Omega$=20-24~MeV, where SCGF-ADC($n$) better reproduces the empirical charge radii and density distribution.

\subsection{Phenomenological Considerations on Occupations, Spectroscopic Factors and the Bubble Structure}

\begin{figure}[h!]%
\includegraphics[width=.5\textwidth]{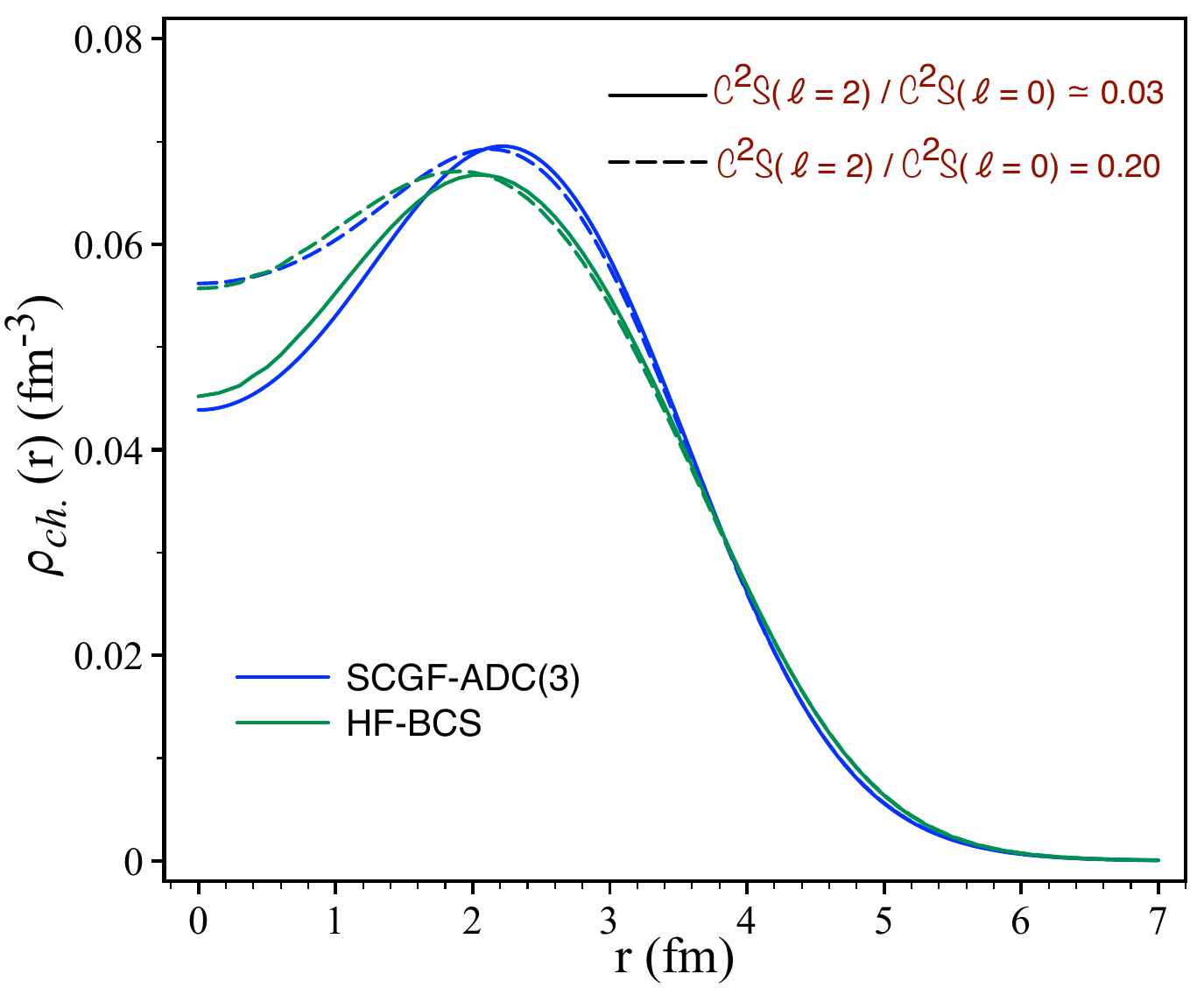}
\caption{Impact of an artificially increased ratio of spectroscopic factors on on the charge distribution of $^{46}$Ar. Full lines refer to ratios close to the \emph{ab initio} predictions, while dashed lines are phenomenological corrections to generate a ratio equal to the experimental value plus one standard deviation.
The SCGF are based on the ADC(3) truncation and the NNLOsat Hamiltonian. The HF-BCS exploit the SLy5-Tw Skyrme functional plus a tunable pairing term.}
\label{fig:1sigma_distr}
\end{figure}
Present measurements of charge distributions are only feasible for stable nuclei. Hence, any study of a bubble structure in $^{46}$Ar can only be indirect for the time being. This work provides evidence of the phenomenon by combining the experimental finding for an empty s$_{1/2}$ proton orbital with state-of-the-art ab initio predictions. The emptiness of this state is a fundamental mechanism for a charge bubble. \emph{All} ab initio simulations that are capable of reproducing the known data on radii also predict an almost complete depletion of the s$_{1/2}$ and consequently a very marked charge bubble in the middle of $^{46}$Ar.
There remains an open question as to whether the experimental uncertainties may allow for a partial s$_{1/2}$ occupation and whether this could weaken the case for a charge bubble. 

We study the correlation between the occupation of the orbitals and the bubble structure through phenomenological corrections on both to \emph{ab initio} and DFT simulations.
We modify the SCGF-ADC(3) spectral function by promoting approximately 0.36 protons from the main d$_{3/2}$ quasihole fragment of $^{45}$Cl to the corresponding s$_{1/2}$ one. As a consequence, the same amount of protons must be shifted from the dominant s$_{1/2}$ to the d$_{3/2}$ quasiparticle fragment in $^{47}$K. This shift results in a new ratio of spectroscopic factors of ${\cal C}^2{\cal S}(\ell=2)/{\cal C}^2{\cal S}(\ell=0)\approx$~0.20 which corresponds to a 1-$\sigma$ deviation in experiment (see Fig.~\ref{fig:lh}). The corresponding change in the charge profile is displayed in Fig.~\ref{fig:1sigma_distr} and shows a halving of the central depletion but it does not eliminate the bubble structure. The reduction of the central depletion varies from $F$=0.37 to 0.19.

Analogous simulations were performed within the standard HF-BCS approach using the Skyrme functional SLy5-Tw, with tensor terms, introduced in Ref.~\cite{Bai2010Sly5Tw5}. The relative occupation of the s$_{1/2}$ to the d$_{3/2}$ can be tuned by varying the surface pairing term. Fig.~\ref{fig:1sigma_distr} shows that the resulting charge distribution is very close to the \emph{ab initio} prediction and its modified spectral function when the same ratios of factors are reproduced, respectively at the experimental central value and its 1-$\sigma$ deviation. 

We stress that all our \emph{ab initio} simulations do not have tunable parameters and they consistently predict a strong charge bubble, unless the spectral function is modified \emph{on purpose} as discussed above. In both microscopic SCGF and HF-BCS simulations, plausible variations of the occupancy with respect to our central experimental value, would require imposing \emph{ad hoc} corrections that weaken but do not entirely eliminate the bubble structure.


\end{document}